# Comparing Imputation Methods for Clinical Prediction Model Development under Complex Missingness Scenarios: A Simulation Study Using Real-World Cardiac Data

Pakpoom Wongyikul[1], Noraworn Jirattikanwong[1], Natthanaphop Isaradech[2], Wuttipat Kiratipaisarl[2], Arintaya Phrommintikul[3], Wachiranun Sirikul[2], Phichayut Phinyo[1]

[1] Department of Biomedical Informatics and Clinical Epidemiology (BioCE), Faculty of Medicine, Chiang Mai University, Chiang Mai, Thailand.

[2] Department of Community Medicine, Faculty of Medicine, Chiang Mai University, Chiang Mai, Thailand.

[3] Division of Cardiology, Department of Internal Medicine, Faculty of Medicine, Chiang Mai University, Chiang Mai, Thailand.

**Correspondence:** Phichayut Phinyo; Department of Biomedical Informatics and Clinical Epidemiology (BioCE), Faculty of Medicine, Chiang Mai University, Chiang Mai, Thailand. Email: phichayutphinyo@gmail.com

## Abstract

**Purpose:** Evidence remains limited on how the imputation methods influence prediction stability across different predictor–outcome relationships and level of missingness. This study evaluates how imputation methods influence overall predictive performance, prediction stability, and computational efficiency across different missingness scenarios.

**Patient and Methods:** A simulation study was conducted using a fully observed real-world cardiac cohort dataset of 8,245 patients, split equally into development and external validation cohorts. Missing data were artificially induced under a missing-at-random mechanism across 18 scenarios, varying by variable type, predictor–outcome relationship, and missingness proportion. Five imputation methods were compared: complete case analysis (CCA), Multiple Imputation by Chained Equations (MICE) with Fully Conditional Specification (MICE-FCS), MICE with predictive mean matching (MICE-PMM), missForest, and k-Nearest neighbors (kNN). Logistic regression models were developed using backward stepwise elimination. Outcomes included optimism-corrected AUC, calibration slope, mean absolute prediction error, external validation performance, and computation time.

**Results:** When missingness involved only isolated linear or categorical variables at 30%–60%, all methods maintained discrimination comparable to the complete-data model, with median AUCs around 0.75. In contrast, when missingness involved isolated non-linear or more complex missing patterns, both predictive performance and calibration worsened as missingness increased, particularly at 90%. In these complex scenarios, MICE showed increased prediction instability and overfitting, while missForest performed well internally but overfitted substantially in external validation. kNN showed the most consistent performance, with more stable predictions, better external validation results, and the shortest computation time across scenarios.

**Conclusion:** Optimal imputation method for clinical prediction modeling may depend on the characteristics of the missing variables. Although MICE preserved predictive performance and CPM reliability in many scenarios, its performance deteriorated in complex settings involving non-linear variables and extreme missingness (90%). For sufficiently large development samples, kNN may provide a computationally efficient alternative imputation strategy for clinical prediction model development.

# Introduction

Clinical prediction models (CPMs) estimate the probability of health-related outcomes for individual patients. By informing diagnosis, prognosis, and treatment selection, they serve as essential tools for patient stratification and tailored clinical management [1]. To fulfill this potential in precision medicine, CPMs must be constructed using rigorous developmental methodology[2]. However, this process is frequently compromised by missing predictor data[2, 3, 4]. If there is a significant proportion of missing data, addressing this problem using complete case analysis may result in a severe reduction in the effective sample size, causing a loss of statistical power and precision[4]. Consequently, complete case models are highly susceptible to bias and poor predictive performance, as the reduced dataset frequently fails to capture the true prognostic weight of important clinical predictors [4].

While descriptive and causal studies prioritize unbiased parameter estimation, clinical prediction research primarily demands optimized predictive accuracy and reliability for future patients. Consequently, the methodological considerations for handling missing data are distinct from those in descriptive or causal studies [5]. Although standard metrics like discrimination and calibration are essential for evaluating CPM predictive performance, they do not guarantee reproducible individual-level predictions [6]. Prediction stability, a complement to these standard performance metrics, is increasingly recognized as a key indicator of model reliability [6, 7]. Prediction instability occurs when a CPM produces significantly different risk estimates for the same individual when the model is developed on different samples drawn from the same target population [6]. Several factors can affect the stability of model performance, including development sample size and heterogeneity, the number and distribution of candidate predictors, outcome prevalence, model complexity [8, 9]. Recent work has shown that the choice of imputation method during development may also influence this prediction stability [7].

Various methods have been proposed to impute missing values prior to modelling. Historically, Multiple Imputation by Chained Equations (MICE) has been the standard approach for handling missing data, as it accounts for estimate uncertainty by generating multiple datasets [10, 11]. A classic MICE fits iterative, parametric imputation models for each missing variable, one by one using Fully Conditional Specification (FCS) [11]. Since clinical data rarely follows strict linear constraints, the rigid assumptions of standard MICE-FCS may lead to model misspecification [12]. To address this, predictive mean matching (PMM) is often integrated into the MICE framework. By drawing imputations directly from observed donor values, MICE-PMM preserves natural estimate variance and prevents extrapolation of imputed value [12, 13]. Alternatively, *missForest*, a non-parametric algorithm based on an optimized random forest, capable of capturing complex, non-linear interactions [14, 15]. As a deterministic single-imputation approach, *missForest* sacrifices the estimation of uncertainty in exchange for a simpler imputed dataset. Although simulation evidence indicates that *missForest* and MICE provide comparable discrimination and stability, *missForest* is more susceptible to overfitting [7]. k-Nearest Neighbors (kNN) presents a pragmatic alternative[16]. By imputing missing values using the median or mode of the most similar cases, the less complex mechanism with single imputation approach of kNN may mitigate overfitting and provide computational efficiency[15].

Crucially, the ability to handle non-linear relationships and interactions during imputation remains vital for CPM development [7, 17]. However, evidence remains limited on how the imputation methods influence prediction stability across different predictor–outcome relationships, particularly when considering their interaction with the proportion of missing data. Method that omitting the stochastic residual draw during development might generate artificial precision and compromises true model internal validation, even with a large sample size [7]. This study conducts a comprehensive scenarios-base study using sufficiently large real world dataset under missing at random (MAR) to compare multiple imputation approaches (MICE-FCS, MICE-PMM) with single imputation methods (*missForest*, kNN). We evaluate these methods across the full model development process, rigorously assessing the integrity of imputed values, variable selection and linear predictor stability, overall predictive performance, prediction stability, and computational efficiency.

# Methods

## Study data and variables

A fully observed real-world clinical dataset comprising 8,245 of cardiac cohort patients and 15 variables (excluding transformed variables). The dataset was partitioned into two distinct cohorts: a development cohort (50%) and an external validation cohort (50%) [**Figure 1**]. Data partitioning was performed using stratified random sampling, based on the primary cardiac outcome, to ensure balanced event rates between the development and validation datasets. The dataset included both continuous clinical measurements (e.g., age, estimated glomerular filtration rate [eGFR], diastolic blood pressure [DBP], high-density lipoprotein [HDL], and non-HDL cholesterol) and categorical patient characteristics (e.g., sex, diabetes mellitus, hypertension, chronic kidney disease, smoking status, and antiplatelet use). Characteristics of patient development and external validation cohort are shown in **Supplementary Table S1.**

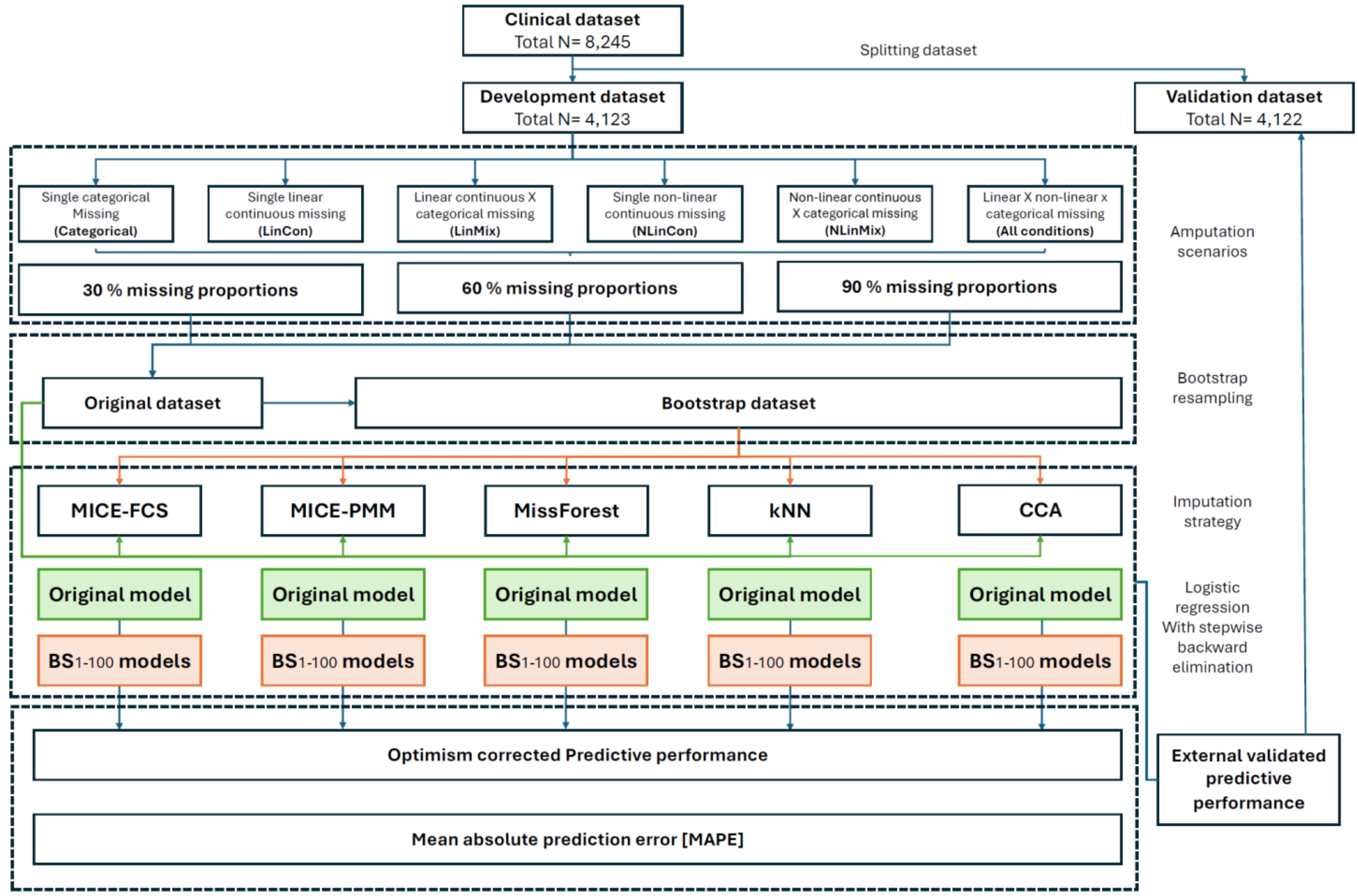

**Figure 1. Study simulation flow diagram.** The figure summarizes the full analytical pipeline of the simulation study. After splitting datasets into development and validation datasets. The upper section illustrates the generation of missing data using different amputation scenarios and missingness proportions. The middle sections show the model development workflow, including bootstrap resampling, application of each imputation strategy, logistic regression modelling with variable selection. The last section involves estimation of optimism-corrected performance, MAPE, and external validation performance. **Abbreviation:** BS, bootstrap sample; CCA, complete-case analysis; kNN, k-nearest neighbors; MAPE, mean absolute prediction error; MICE-FCS, multiple imputation by chained equations using fully conditional specification; MICE-PMM, multiple imputation by chained equations using predictive mean matching.

**Sample size calculation**

The required minimum sample size for model development was estimated as 1,647 participants [18], based on an expected AUC of 0.75, 18 candidate predictors, and an outcome prevalence of 0.13. To avoid the influence of small sample size masking the intended isolated differences between imputation methods, we used a fixed initial sample size of 4,123, approximately 2.5 folds to the minimum requirement.

**Missing data simulation**

To systematically evaluate the impact of missing data, we artificially induced missingness into the fully observed dataset using multivariate amputation, implemented via the ampute function in R [19]. All missing data were generated strictly under MAR mechanism, wherein the probability of a value being missing is systematically related to the observed patient characteristics, but not to the unobserved values themselves.

We constructed 18 distinct simulation scenarios by manipulating two core dimensions:

1. **Target variables and types:** Missingness was imposed on a single categorical variable (antiplatelet use), a single continuous variable (HDL or eGFR), or a mixture of continuous and categorical variables. The probability of missingness was modeled to depend on other fully observed variables through different specified weights, simulating linear and non-linear relationships. Missingness was heavily weighted by specific target variables as shown in **Supplementary Table S2.**
2. **Missingness proportions:** Each missingness pattern was generated at a moderate (30%), a high (60%), and an extremely high (90%) proportion of missing data.

**Post-Amputation data processing**

Clinical prediction models frequently incorporate non-linear and interaction effects to better capture realistic and clinically meaningful associations between predictors and outcomes. Therefore, after the missing data generation was completed for each scenario, we derived quadratic terms for age, eGFR, and DBP ($age^2$, $eGFR^2$, $DBP^2$) and interaction terms between age and renal/lipid profiles (age × eGFR, age × non-HDL). In total, 18 candidate predictors were included in initial model. Generating these derived variables after amputation mirrors the real-world clinical scenario where complex terms must be handled in the presence of missing baseline data.

**Table 1. Missing data simulation scenarios**

| Scenario | Proportion missing | Variable(s) with missing data | Variable type | Dependency / relationship to outcome | Complete cases remaining |
|---|---|---|---|---|---|
| **1** | 30% | Antiplatelet | Single Categorical | Categorical | 2,915 |
| **2** | 60% | Antiplatelet | Single Categorical | Categorical | 1,664 |
| **3** | 90% | Antiplatelet | Single Categorical | Categorical | 397 |
| **4** | 30% | HDL | Single Continuous | Linear relationship (LinCon) | 2,915 |
| **5** | 60% | HDL | Single Continuous | Linear relationship (LinCon) | 1,729 |
| **6** | 90% | HDL | Single Continuous | Linear relationship (LinCon) | 433 |
| **7** | 30% | eGFR | Single Continuous | Non-linear relationship (NLinCon) | 2,924 |
| **8** | 60% | eGFR | Single Continuous | Non-linear relationship (NLinCon) | 1,679 |
| **9** | 90% | eGFR | Single Continuous | Non-linear relationship (NLinCon) | 404 |
| **10** | 30% | Antiplatelet, HDL | Mixed (Cat. + Cont.) | Linear relationship (LinMix) | 2,833 |
| **11** | 60% | Antiplatelet, HDL | Mixed (Cat. + Cont.) | Linear relationship (LinMix) | 1,652 |
| **12** | 90% | Antiplatelet, HDL | Mixed (Cat. + Cont.) | Linear relationship (LinMix) | 382 |

| Scenario | Proportion missing | Variable(s) with missing data | Variable type | Dependency / relationship to outcome | Complete cases remaining |
|---|---|---|---|---|---|
| **13** | 30% | Antiplatelet, eGFR | Mixed (Cat. + Cont.) | Non-linear relationship (NLinMix) | 2,850 |
| **14** | 60% | Antiplatelet, eGFR | Mixed (Cat. + Cont.) | Non-linear relationship (NLinMix) | 1,660 |
| **15** | 90% | Antiplatelet, eGFR | Mixed (Cat. + Cont.) | Non-linear relationship (NLinMix) | 410 |
| **16** | 30% | Antiplatelet, HDL, eGFR | Mixed (All) | All conditions | 2,900 |
| **17** | 60% | Antiplatelet, HDL, eGFR | Mixed (All) | All conditions | 1,635 |
| **18** | 90% | Antiplatelet, HDL, eGFR | Mixed (All) | All conditions | 383 |

**Abbreviation:** Cat., categorical; Cont., continuous; eGFR, estimated glomerular filtration rate; HDL, high-density lipoprotein cholesterol; LinCon, linear relationship for a single continuous missing variable; LinMix, linear relationship for mixed categorical and continuous missing variables; NLinCon, non-linear relationship for a single continuous missing variable; NLinMix, non-linear relationship for mixed categorical and continuous missing variables.

**Handling of missing data**

To systematically evaluate the impact of missing data handling on the development and stability of the CPM, we implemented and compared five distinct imputation methods. Complete Case Analysis (CCA), where records with missing values were omitted; MICE-FCS, which used iterative parametric regression (10 datasets, 5 iterations) and the "Just Another Variable" (JAV) approach for interactions [20]; MICE-PMM, applying predictive mean matching with 10 donors to handle non-normality (10 datasets, 5 iterations) [12, 13]; missForest, a non-parametric random forest imputation (50 trees, 10 iterations) [7, 15, 21]; and kNN imputation (k=10), using the median or mode of neighbors for continuous or categorical variables, respectively [16]. For all methods, the primary outcome was excluded from imputation to prevent data leakage.

**Table 2. Comparison of imputation methods**

| Strategy | Type of imputed dataset | Handling of categorical/continuous variables | Handling of derived terms (non-linear/interactions) |
|---|---|---|---|
| **MICE-FCS** | Multiple | Logistic regression (Cat), Linear regression (Cont) | Just Another Variable (JAV) with restricted predictor matrix |
| **MICE-PMM** | Multiple | Predictive Mean Matching for all types (10 donors) | Just Another Variable (JAV) with restricted predictor matrix |
| **missForest** | Single | Averaging over 50 unpruned classification/regression trees | Treated as independent variables |
| **kNN** | Single | Mode (Cat), Median (Cont) based on k=10 neighbors | Treated as independent variables |

**Abbreviation:** Cat., categorical; Cont., continuous; kNN, k-nearest neighbors; MAPE, mean absolute prediction error; MICE-FCS, multiple imputation by chained equations using fully conditional specification; MICE-PMM, multiple imputation by chained equations using predictive mean matching.

**Model development and internal validation**

To assess model stability and optimism, internal validation was performed using bootstrap resampling [**Figure 1**]. For each simulation scenario and imputation strategy, models were fitted on bootstrap samples drawn with replacement from the original cohort. For the prediction of the cardiac outcome, multivariable logistic regression models were constructed. The maximal (full) model included all predictors alongside the pre-specified quadratic terms ($age^2$, $eGFR^2$, $DBP^2$) and interaction terms (age × eGFR, age × non-HDL).

Subsequent variable selection was performed using stepwise backward elimination governed by a strict P-value threshold of 0.05. For the MICE-imputed datasets, backward elimination was performed across the multiply imputed datasets using Rubin's rules for pooled parameter estimation via the psfmi_lr package [22, 23]. The final selected variables and their pooled coefficients were extracted to formulate the final prognostic models for each bootstrap iteration.

**Model evaluation**

Several dimensions of model performance, prediction stability, and computational efficiency were assessed across the 100 bootstrap iterations for each of the 18 simulation scenarios. The following core metrics were extracted and aggregated:

1. **Integrity of imputed values**

   To deeply understand how each method handled continuous clinical predictors (specifically HDL and eGFR, along with their derived polynomial and interaction terms ($eGFR^2$ and age × eGFR), we extracted the standard deviation (SD) and median of the imputed values from each bootstrap. This allowed us to determine whether certain imputation algorithms artificially distorted or narrowed the natural clinical variance of the biomarkers.

2. **Model structure stability**

   Because imputed value from each imputation method may alter the variance and covariance structures of the dataset, they inherently affect automated variable selection. We evaluated model structure stability by recording the absolute number of clinical predictors successfully retained in the final multivariable models following the backward stepwise elimination procedure ($P < 0.05$). To evaluate model-structure consistency, we fitted each bootstrap-derived model to the original dataset and extracted the distribution of the linear predictor (LP). Specifically, we calculated the mean and standard deviation (SD) of the LP across all bootstrap iterations.

3. **Overall predictive performance**

   Model performance was quantified using optimism-corrected estimates. Optimism was defined as the average difference between model performance in the bootstrap samples and performance of the same bootstrap-derived models when applied to the original dataset. Discrimination was assessed using the optimism-corrected area under the receiver operating characteristic curve (AUC), which reflects the model's ability to distinguish between patients with and without the primary cardiac outcome.

   Calibration was assessed using the optimism-corrected calibration slope, which evaluates the spread of estimated risks. A perfectly calibrated model has a slope of 1.0; values <1.0 indicate overfitting, where predictions are too extreme, whereas values >1.0 indicate underfitting. A deviation of ±0.1 from 1.0 was considered significant miscalibration.

4. **Individual prediction stability and external validation performance**

   Prediction instability was calculated as the absolute difference between the predicted probability from the bootstrap model and that from the original development model. These absolute differences were summarized using the mean absolute prediction error (MAPE) [6]. We also calculated the MAPE in percentage by dividing the MAPE with predicted probability from original development model. Since bootstrap internal validation can still yield optimistic performance estimates by its resampling nature as approximately 63.2% of the original sample is typically included in each bootstrap sample [24]. Therefore, we validated the derived CPM in an external dataset, assessing both discrimination and calibration to evaluate potential overfitting.

5. **Computational efficiency**

   To inform practical guidelines for clinical researchers, we logged the computational time (in seconds) required to execute the full pipeline comprising: bootstrap resampling, missing data imputation, derived term calculation, and final model fitting, for each strategy.

**Software and computational resources**

The simulation study was conducted using the R Project for Statistical Computing software (version 4.5.2) and performed using the high-performance computing (HPC) resources of the Faculty of Medicine, 161 Chiang Mai University, including the MedCMU-HPC-RAPTOR cluster.

## Results

### 1. Integrity of imputed values

Across the 30%, 60%, and 90% missingness scenarios, the MICE approaches (MICE-FCS and especially MICE-PMM) preserved the variability of continuous predictors, both linear and non-linear, most closely to the complete-data reference, with SDs remaining largely comparable [**Supplementary Table S3**]. MICE-FCS showed widening SDs as the proportion of missingness increased. In contrast, kNN maintained variance well only at 30% missingness for non-linear predictors. Variance progressively reduced as missingness increased, with this shrinkage more pronounced for linear predictors. missForest consistently underestimated variance across all scenarios. Regarding central tendency, MICE-PMM, produced medians closest to the complete-data reference for both linear and non-linear predictors [**Supplementary Table S4**]. Among the machine-learning methods, missForest generally maintained medians comparable to MICE-FCS and CCA, whereas kNN tended to overestimate imputed values, particularly for non-linear predictors across all missingness proportions.

### 2. Model structure stability

After backward stepwise elimination, all imputation methods retained a similar number of predictors (approximately 7–10). MICE-PMM and MICE-FCS retained the largest number of clinical predictors in most scenarios, except when missingness reached 90% under non-linear missing scenarios [**Figure 2**]. Similarly, the mean and SD of the linear predictor (LP) were similar across most scenarios; however, at 90% missingness in the LinMix, NLinCon, NLinMix, and All conditions, MICE, particularly MICE-FCS, produced a notably high variance of mean and SD of the LP [**Figure 3Aand 3B**]. Detail on number of retained predictor and LP are provided in **Supplementary Table S5**

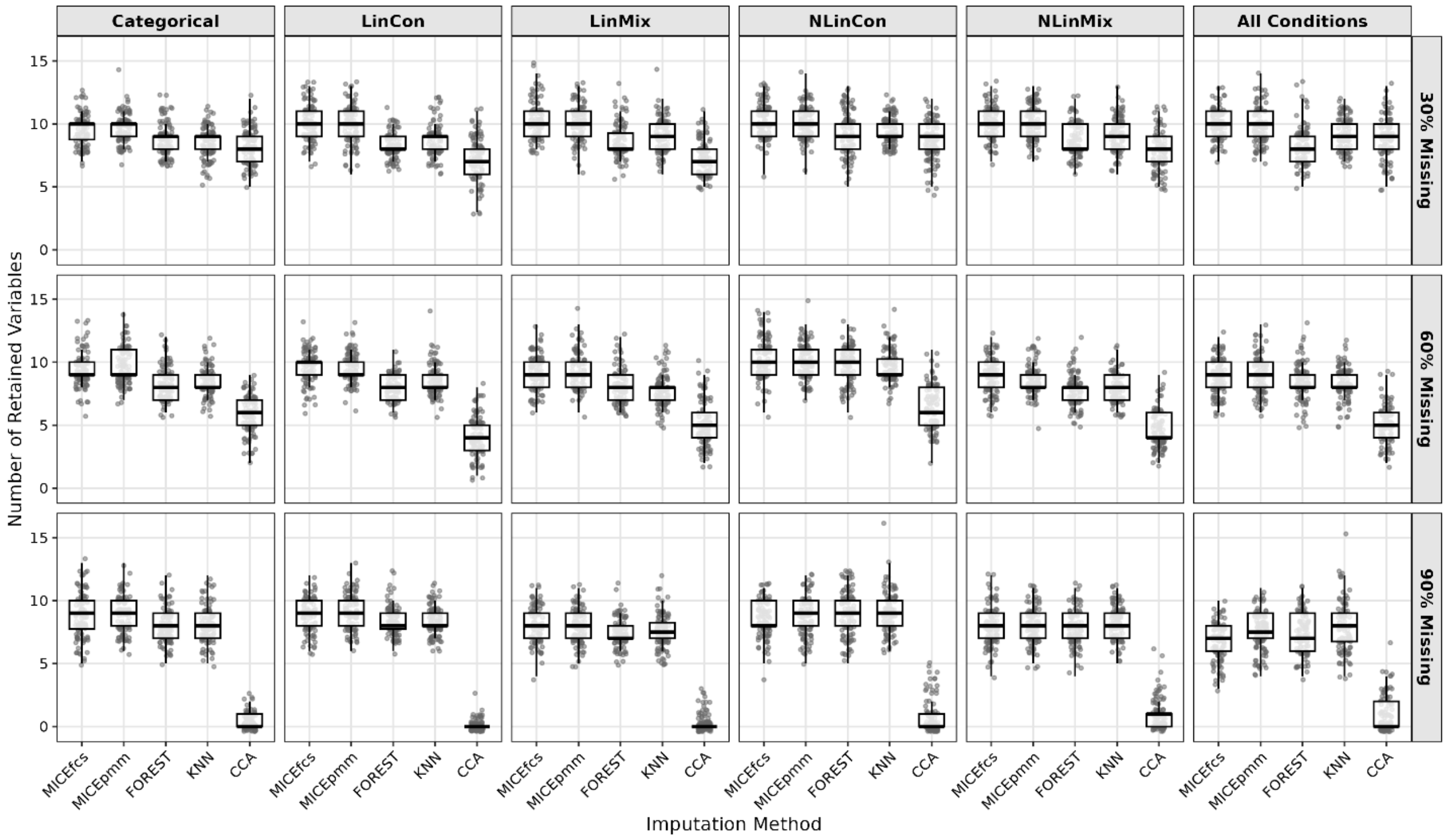

**Figure 2. Comparison of number of retained predictors across all scenarios and imputation methods**. Boxplots show the distribution of the number of retained predictors after model selection across imputation methods, missing-data scenarios, and missingness proportions. **Abbreviations:** CCA, complete-case analysis; FOREST, missForest imputation; kNN, k-nearest neighbors; MAPE, mean absolute prediction error; MICE-FCS, multiple imputation by chained equations using fully conditional specification; MICE-PMM, multiple imputation by chained equations using predictive mean matching.

(A)

(B)

**Figure 3. Comparison of standard deviation (A) and median (B) of linear predictors (LP) across all scenarios and imputation methods. Abbreviation** CCA, complete-case analysis; FOREST, missForest imputation; kNN, k-nearest neighbors; MAPE, mean absolute prediction error; MICE-FCS, multiple imputation by chained equations using fully conditional specification; MICE-PMM, multiple imputation by chained equations using predictive mean matching; LP, linear predictors

### 3. Overall predictive performance

When missingness occurred in Categorical, LinCon, and LinMix settings, all imputation methods preserved optimism-corrected discrimination comparable to the complete-data model (median AUC ≈ 0.75) across missingness levels. The machine-learning methods showed a slight reduction in AUC (kNN: 0.74 [IQR 0.73–0.75]; missForest: 0.73 [IQR 0.72–0.74]). In contrast, under NLinCon, NLinMix, and All conditions, AUC decreased progressively as the proportion of missing data increased [**Figure 4A**]. Calibration slopes were generally maintained within the acceptable range (0.9–1.1) across most data types and missingness levels [**Figure 4B**]. However, at 90% missingness in the NLinCon and All conditions, MICE methods showed a notable decline in calibration slope (<0.9), whereas the machine-learning approaches remained within the optimal range. Details on predictive performance is provided in **Supplementary Table S6**

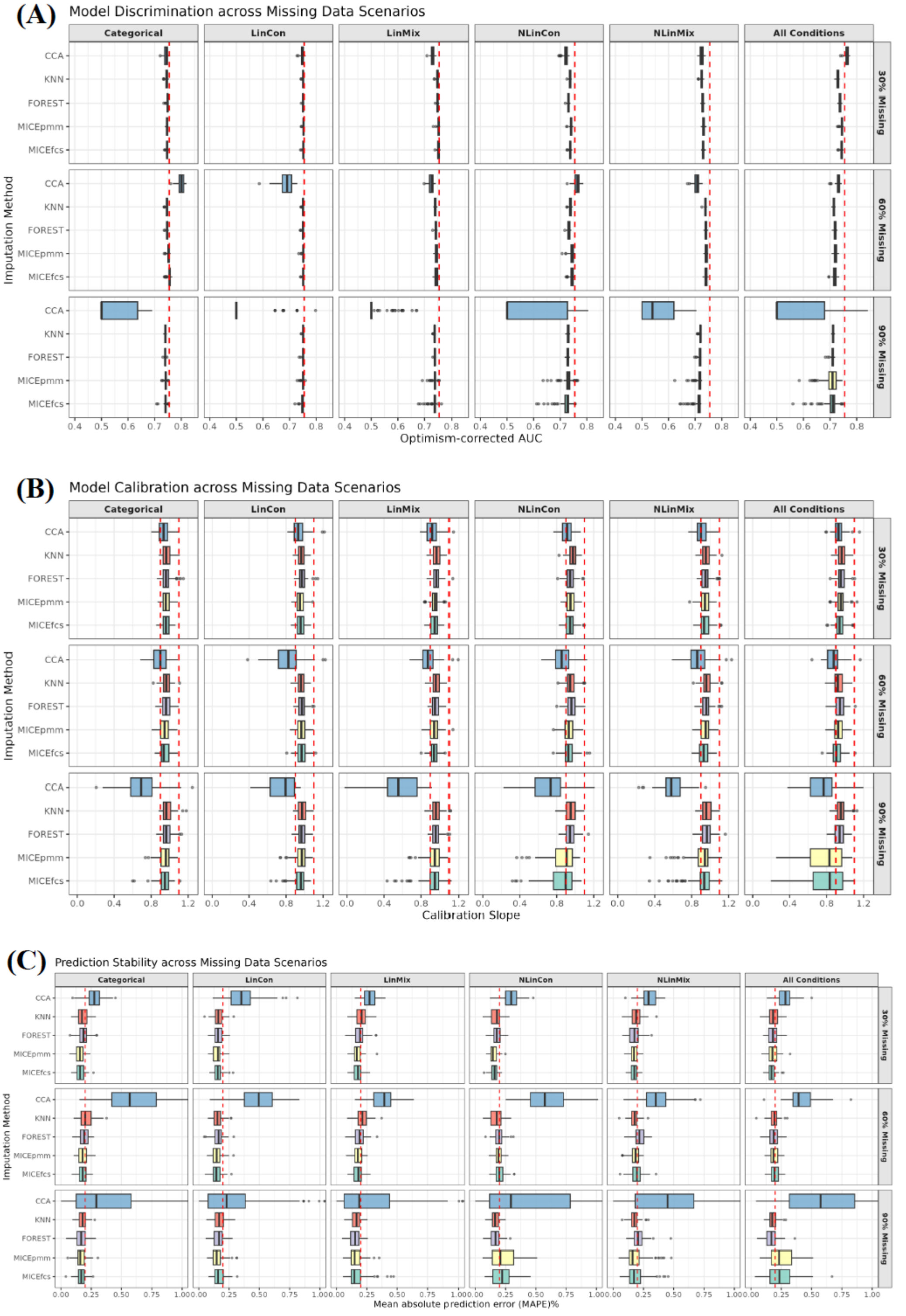

**Figure 4 Comparison of optimism-corrected AUC, calibration slope, and mean absolute prediction error (MAPE) across all scenarios and imputation methods.** Panel (A) shows optimism-corrected discrimination, assessed by the area under the receiver operating characteristic curve (AUC). Red dashed vertical line indicates reference performance at 0.75 based on full dataset model. Panel (B) shows optimism-corrected calibration slope. Red dashed vertical lines range optimal calibration threshold. Panel (C) shows prediction stability, assessed using mean absolute prediction error (MAPE%) Red dashed vertical line indicates optimal MAPE % threshold at 20%. Results are presented across missing data patterns (Categorical,

LinCon, LinMix, NLinCon, NLinMix, and All Conditions), missingness proportions (30%, 60%, and 90%).

**Abbreviation** CCA, complete-case analysis; FOREST, missForest imputation; kNN, k-nearest neighbors; MAPE, mean absolute prediction error; MICE-FCS, multiple imputation by chained equations using fully conditional specification; MICE-PMM, multiple imputation by chained equations using predictive mean matching

**4. Prediction stability and external validation performance**

MICE-FCS and MICE-PMM generally achieved slightly lower and more stable MAPE in the Categorical, LinCon, and LinMix settings at 30% and 60% missingness [**Figure 4C**]. In contrast, missForest and kNN performed competitively at moderate to high missingness when non-linear variables were missing, and showed greater stability and even lower MAPE at 90% missingness [**Supplementary Table S6**]. Instability plots and MAPE plots across all scenarios are provided in **Supplement Figure S1-S5**.

Predictive performance in external validation indicated that, under the Categorical, LinCon, and LinMix scenarios, all methods largely preserved both discrimination and calibration, with the MICE approaches showing the best calibration across missingness levels [**Figure 5**]. However, in scenarios involving non-linear missingness, missForest showed marked overfitting. MICE also exhibited increasing degree of overfitting as the proportion of missing data rose. In contrast, kNN maintained comparatively stable performance, with suboptimal discrimination and calibration across these non-linear scenarios. Details on external validation predictive performances are provided in **Supplementary Table S7.**

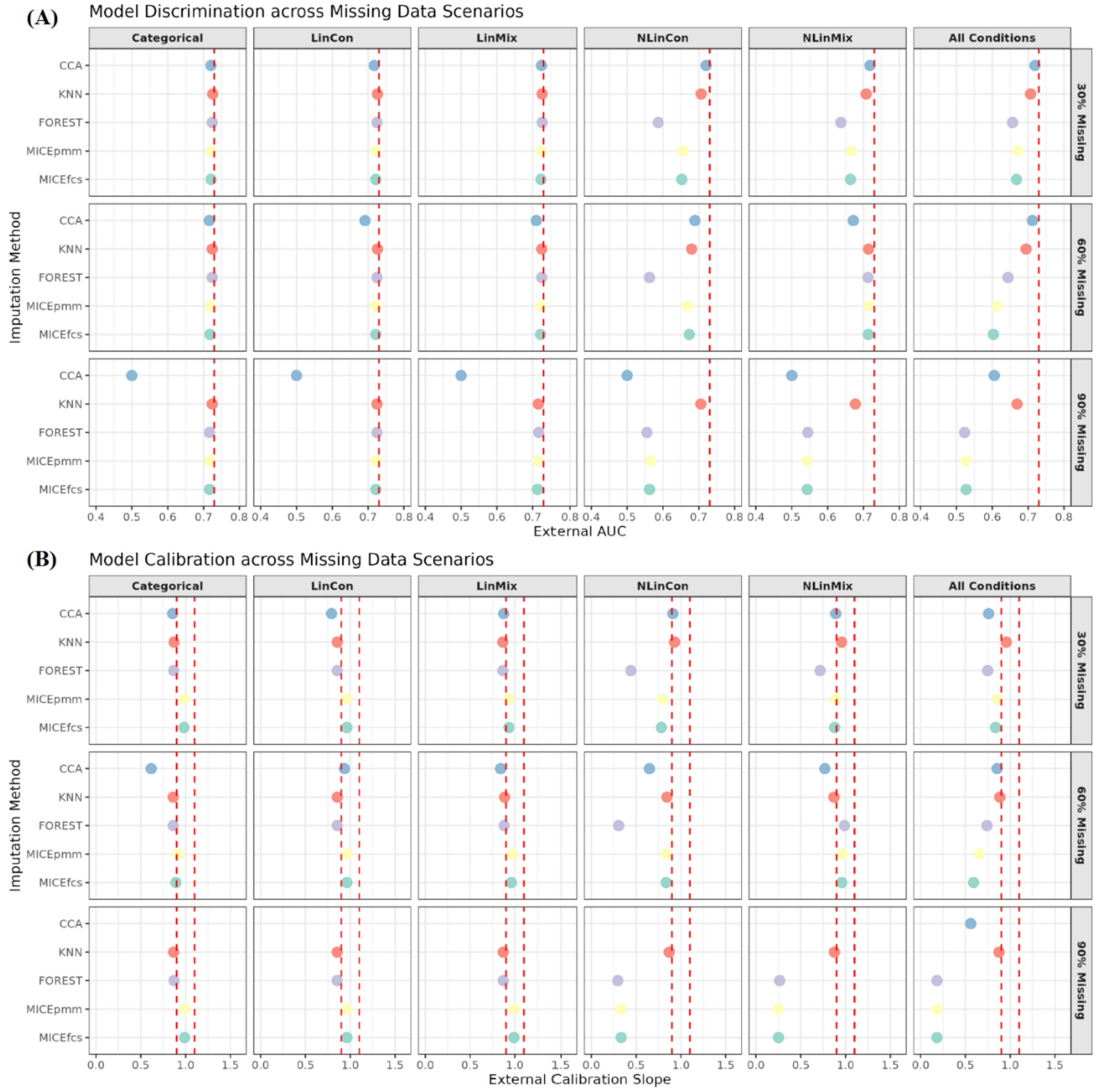

**Figure 5. Comparison of AUC, calibration slope across all scenarios and imputation methods based on external validation dataset.** Panel (A) shows external validated discrimination, assessed by the area under the receiver operating characteristic curve (AUC). Red dashed vertical line indicates reference performance at 0.71 based on full dataset model. Panel (B) shows external validated calibration slope. Red dashed vertical lines range optimal calibration threshold. **Abbreviation** CCA, complete-case analysis; FOREST, missForest imputation; kNN, k-nearest neighbors; MAPE, mean absolute prediction error; MICE-FCS, multiple imputation by chained equations using fully conditional specification; MICE-PMM, multiple imputation by chained equations using predictive mean matching

**5. Computational efficiency**

Execution time (in seconds) for the full pipeline varied by method. kNN maintained high efficiency across scenarios (e.g., 10.2 seconds [IQR: 9.9, 10.5] in the 60% LinCon scenario) [**Supplementary Figure S2** and **Supplementary Table S6**]. MICE-FCS and MICE-PMM required moderate and consistent computation times, approximately ranging between 61 and 73 seconds across most iterations. The execution time for missForest was highly variable; it required 26.3 seconds (IQR: 10.1, 85.1) in the 60% LinCon scenario but increased significantly under mixed complexities, reaching 78.4 seconds (IQR: 49.3, 93.6) in the 30% "All Conditions" scenario.

## Discussion

Based on real-world clinical data, this study evaluated how four common imputation methods and a complete-case analysis approach influence CPM development, spanning imputed-value fidelity, model-structure stability, prediction stability, predictive performance (internal and external) and computational efficiency. Although MICE generally preserved the integrity of imputed values and model-structure stability across most data types at moderate to high missingness, its clear advantage diminished in complex non-linear setting, where performance became comparable to single-imputation methods and was sometimes worse at 90% missingness. A similar pattern was observed at the individual prediction level, with MICE becoming notably less stable under non-linear scenarios at 90% missingness. missForest appeared acceptable in terms of model structure stability, predictive performance, and individual-level predictions in internal validation; however, it exhibited significant overfitting in external validation under non-linear setting, with degree of overfitting worsening as missingness increased. In contrast, despite producing biased imputation with progressive variance shrinkage as also observed in missForest, kNN showed greater stability from model structure to individual prediction in both internal and external validation and outperformed MICE in certain non-linear scenarios.

A recent simulation study specifically evaluating missing data at the point of deployment suggested that a model with unbiased model parameters is not necessarily the one that predicts best, particularly when handling missing data at the point of care [10, 17, 25]. Consequently, when developing CPM intended for real-world application, the recovery of perfectly unbiased parameter estimates is no longer the primary concern [17]. In our study, the MICE approach generally preserved the integrity of imputed values, with only minimal deterioration under non-linear scenarios and no clear superiority compared with machine-learning methods. However, under extremely high missingness, the parametric components within MICE appeared insufficient to maintain model-structure stability, predictive performance, and individual-level prediction stability in non-linear scenarios. Although MICE-PMM is designed to better accommodate non-linear relationships and reduce extrapolation[12], its semi-parametric regression structure becomes unstable when the proportion of completely observed rows shrinks to 10%.

Based on internal validation, CPMs developed using missForest and kNN imputation showed better predictive performance and prediction stability under extremely high missingness of non-linear scenarios. One possible explanation is their non-parametric nature, which makes them less dependent on strict model-specification assumptions and more suitable for handling complex predictor–outcome relationships [7, 15]. In addition, unlike MICE, these methods do not rely on stochastic draws from posterior predictive distributions and therefore tend to underestimate uncertainty associated with imputed value [13, 17]. In our study, however, when resampling techniques were applied, the reproducibility of case mix, represented by the mean and standard deviation of the LP, was not substantially affected by this variance compression. On the other hand, MICE required imputed values to be generated under parametric assumptions, with additional uncertainty introduced through stochastic sampling [11]. The result based on our dataset environment suggested that when these assumptions are violated, particularly in complex non-linear settings with very high missingness, the combination of model misspecification, stochastic imputation, and bootstrap internal validation may contribute to reduced predictive performance and increased prediction instability.

In our study, backward stepwise elimination was used to reduce the risk of prematurely excluding potentially important predictors. For predictors with extremely high missingness, imputation might allowed them to enter the model development process, after which they could be retained or removed through automated predictor selection. If such predictors remain in the final model and the CPM shows acceptable internal validation performance but low prediction stability, external validation remains the reference standard for evaluating CPM performance and transportability. In our external validation, missForest showed substantial overfitting under non-linear scenarios, followed by MICE across missingness levels. Surprisingly, kNN, despite being a less complex machine-learning method, could handle low dimensional data and preserved optimal predictive performance under the same scenarios. These findings suggest that introducing more complex imputation methods during model development may reduce bias in some settings but may also increase model variance, leading to poorer transportability and overfitting [7, 26].

At 30%–60% missingness with sufficiently large sample size, no single imputation method clearly outperformed the others. While our study suggests that kNN imputation may provide promising performance compared with other methods at 90% missingness, some may argue that such extreme missingness reflects limited predictor availability and predictors with such high missingness might reasonably be excluded during the initial model development stage.[27, 28]. This is partly true; however, in some contexts, it may not be appropriate. In an era of rapidly evolving medical innovation and changing healthcare workflows, newly introduced but clinically well-justified predictors may still be valuable for CPM development [29, 30]. In clinical practice, missing data often occurs at the point of prediction [10, 17, 25, 31]. Therefore, computational efficiency is an important consideration during model deployment. MICE, which generates multiple imputed datasets, is computationally intensive and may be less practical for real-time prediction. In contrast, deterministic imputation methods are straightforward and faster to apply to future patients. However, our result showed that missForest, which incorporates an internal bootstrap loop, may still be computationally demanding in some scenarios. kNN was more computationally efficient, requiring approximately 2–5 times less computation time than MICE.

This study has several strengths and limitations. A key strength is the use of a real-world clinical dataset to compare multiple imputation methods. To provide a comprehensive and interpretable assessment, we evaluated not only predictive performance and individual-level prediction stability within the development dataset, but also imputed-value integrity, model structure based on the number of retained predictors, the distribution of the linear predictor, and performance in an external validation dataset. Several limitations should also be acknowledged. First, despite real clinical practice is used, this study was based on a single dataset; therefore, the observed patterns may not be fully transportable to other datasets. Second, we focused only on MAR scenarios and used a fixed sample size. More complex scenarios, such as combinations of MAR, MCAR, and MNAR mechanisms or varying initial sample sizes, may provide additional information. However, the true missing-data mechanism is generally untestable, and previous evidence suggests that the assumed mechanism may have only a minor impact on predictive performance [7]. We used a fixed initial sample size to avoid masking the intended isolated differences between missingness scenarios and imputation methods. Third, we fitted prediction models using only logistic regression and did not evaluate machine-learning prediction models. Logistic regression was chosen because it is stable, interpretable, and computationally efficient. Finally, prediction stability was assessed using a bootstrap-based MAPE approach with only 100 bootstrap rounds. Therefore, our findings should not be interpreted as confirmatory, but rather as exploratory evidence that contributes to strengthening methodology for CPM development in the presence of missing data.

## Conclusion

This study demonstrated that the optimal imputation method for clinical prediction modeling may depend on the characteristics of the missing variables. Although MICE preserved predictive performance and the reliability of CPMs in many scenarios, its performance deteriorated in complex settings involving non-linear variables and extreme missingness (90%), with increased prediction instability and overfitting. Alternatively, the kNN algorithm maintained superior prediction stability without the marked overfitting seen in missForest. Combined with its computational efficiency advantage, and given that the development sample size is sufficiently large, a simpler deterministic approach such as kNN can offer an alternative imputation choice for clinical prediction model development.

### Abbreviations

**AUC**, Area Under the Receiver Operating Characteristic Curve; **BS**, Bootstrap Sample; **CCA** ,Complete Case Analysis; **CPM**, Clinical Prediction Model; **FCS**, Fully Conditional Specification; **IQR**, Interquartile Range; **JAV**; Just Another Variable; **kNN**, k-Nearest Neighbors; **LP**, Linear Predictor; **MAPE**, Mean Absolute Prediction Error; **MAR**, Missing at Random; **MCAR**, Missing Completely at Random; **MICE**, Multiple Imputation by Chained Equations; **MNAR**, Missing Not at Random

### Acknowledgements

This study was partially supported by Chiang Mai University and the Faculty of Medicine, Chiang Mai University. The authors also acknowledge the MedCMU-HPC-RAPTOR cluster, Faculty of Medicine, Chiang Mai University, for providing high-performance computing (HPC) resources for the bootstrap resampling analyses.

### Author Contributions

PW, NJ, NI, WK, AP, WS, and PP conceived the study. All authors were involved in collection and management of data, conception, study design, and interpretation. PW and PP were involved in data curation, methodology, formal analysis, visualization, writing – original draft; PP took part in drafting, revising or critically reviewing the article; gave final approval of the version to be published.

## Ethics approval

Ethical approval was granted by the Institutional Review Board and Ethics Committee of the Faculty of Medicine, Chiang Mai University (HOS-2568-0283) and was exempted from ethics review as it involved secondary analysis of de-identified data from a previously approved clinical prediction model development study.

## Data availability

All data generated or analyzed during this study are included in this published article and its supplementary information files.

## Funding

This study receives no external funding.

## Conflict of interest

None declared.

## Declaration of Generative AI and AI-assisted technologies in the writing process

During the preparation of this manuscript, the authors used ChatGPT-5 and Gemini to assist with language editing, grammar correction, and clarity. All AI-assisted outputs were reviewed, edited, and verified by the authors, who take full responsibility for the content of the manuscript.

# Supplementary Appendix

**Supplementary Table S1.** Baseline characteristics of initial model predictors in development and validation cohort.

**Supplementary Table S2.** Amputation Patterns and Structural Predictor Configurations under Missing at Random (MAR) Scenarios.

**Supplementary Table S3.** Comparison of the standard deviation of imputed values across missingness scenarios and imputation methods.

**Supplementary Table S4.** Comparison of the median of imputed values across missingness scenarios and imputation methods.

**Supplementary Table S5.** Number of retained predictors in model and distribution of Linear predictor (LP)

**Supplementary Table S6**. Optimism-corrected performance, prediction stability, and computational time.

**Supplementary Table S7.** External validation of predictive performance.

**Supplementary Figure S1.** Instability plot and mean absolute predictor error (MAPE) plot across all missingness proportions in missing of single categorical scenarios

**Supplementary Figure S2.** Instability plot and mean absolute predictor error (MAPE) plot across all missingness proportions in missing of single linear continuous (LinCon) scenarios

**Supplementary Figure S3.** Instability plot and mean absolute predictor error (MAPE) plot across all missingness proportions in missing of linear continuous and categorical (LinMix) scenarios

**Supplementary Figure S4.** Instability plot and mean absolute predictor error (MAPE) plot across all missingness proportions in missing of single non-linear continuous (NLinCon) scenarios

**Supplementary Figure S5.** Instability plot and mean absolute predictor error (MAPE) plot across all missingness proportions in missing of non-linear continuous and categorical (NLinMix) scenarios

**Supplementary Figure S6.** Instability plot and mean absolute predictor error (MAPE) plot across all missingness proportions in missing of all condition's scenarios

**Supplementary Figure S7.** Comparison of Computational time across all scenarios and imputation methods

**Supplementary Table S1.** Baseline characteristics of initial model predictors in development and validation cohort.

| No. | Predictor (Unit) | Type of Data | Development<br>N = 4,123<br>[n (%) or Median (IQR)] | Validation<br>N = 4,122<br>[n (%) or Median (IQR)] |
|---|---|---|---|---|
| 1 | Age (years) | Continuous (Linear) | 66.0 (59.0 – 73.0) | 66.0 (58.0 – 73.0) |
| 2 | Age squared | Continuous (Quadratic) | 4356.0 (3481.0 – 5329.0) | 4356.0 (3364.0 – 5329.0) |
| 3 | Sex | Categorical | | |
| | Male | | 1842 (44.7%) | 2253 (54.7%) |
| | Female | | 2281 (55.3%) | 1869 (45.3%) |
| 4 | Diabetes Mellitus | Categorical | | |
| | No DM | | 1749 (42.4%) | 1741 (42.2%) |
| | DM | | 2374 (57.6%) | 2381 (57.8%) |
| 5 | Hypertension | Categorical | | |
| | No HTN | | 153 (3.7%) | 164 (4.0%) |
| | HTN | | 3970 (96.3%) | 3958 (96.0%) |
| 6 | Chronic Kidney Disease | Categorical | | |
| | No CKD | | 3275 (79.4%) | 3282 (79.6%) |
| | CKD | | 848 (20.6%) | 840 (20.4%) |
| 7 | Smoking Status | Categorical | | |
| | No smoking | | 3911 (94.9%) | 3911 (94.9%) |
| | Smoking | | 212 (5.1%) | 211 (5.1%) |
| 8 | Antiplatelet Use | Categorical | | |
| | Not take antiplatelet | | 1116 (27.1%) | 1090 (26.4%) |
| | Take antiplatelet | | 3007 (72.9%) | 3032 (73.6%) |

| No. | Predictor (Unit) | Type of Data | Development<br>N = 4,123<br>[n (%) or Median (IQR)] | Validation<br>N = 4,122<br>[n (%) or Median (IQR)] |
|---|---|---|---|---|
| 9 | eGFR (mL/min/1.73m²) | Continuous (Linear) | 69.29 (50.28 – 88.07) | 68.13 (49.20 – 86.84) |
| 10 | eGFR squared | Continuous (Quadratic) | 4801.44 (2528.43 – 7755.68) | 4642.00 (2420.49 – 7540.79) |
| 11 | Age × eGFR | Continuous (Interaction) | 4477.11 (3372.16 – 5496.38) | 4375.44 (3297.06 – 5426.46) |
| 12 | Triglycerides (mg/dL) | Continuous (Linear) | 120.0 (89.0 – 162.0) | 119.0 (88.0 – 162.0) |
| 13 | HDL Cholesterol (mg/dL) | Continuous (Linear) | 47.0 (40.0 – 57.0) | 47.0 (40.0 – 57.0) |
| 14 | LDL > 70 mg/dL | Categorical | | |
| | LDL less 70 | | 1014 (24.6%) | 1105 (26.8%) |
| | LDL more 70 | | 3109 (75.4%) | 3017 (73.2%) |
| 15 | Age × Non-HDL | Continuous (Interaction) | 7040.0 (5716.0 – 8753.5) | 6954.5 (5621.8 – 8680.0) |
| 16 | Diastolic BP (mmHg) | Continuous (Linear) | 75.0 (67.0 – 81.0) | 74.0 (67.0 – 80.0) |
| 17 | Diastolic BP squared | Continuous (Quadratic) | 5625.0 (4489.0 – 6561.0) | 5476.0 (4489.0 – 6400.0) |
| 18 | Systolic BP (mmHg) | Continuous (Linear) | 132.0 (120.0 – 144.0) | 130.0 (120.0 – 142.8) |

**Missing Data Simulation Mechanics**

Missing values were artificially introduced into the development cohort using a multivariate amputation procedure (mice::ampute) under a Missing at Random (MAR) mechanism. Amputation configurations were meticulously mapped to replicate three structural variations in data patterns: single categorical missingness (antiplatelet use), single continuous missingness (eGFR or HDL cholesterol), and mixed multi-variable missingness combinations.

- Categorical Predictor Amputation (Antiplatelet Use): Missingness was explicitly driven by the baseline physiological and clinical profile vector across the first eight sequential variables in the dataset. This conditioning pool comprised cardiac_outcome, age, sex, dm (diabetes mellitus), htn (hypertension), sbp (systolic blood pressure), dbp (diastolic blood pressure), and egfr (estimated glomerular filtration rate). Each of these observed variables was assigned an equal structural regression weight of 0.2, ensuring that the probability of missing antiplatelet therapy records was directly dependent on the patient's primary clinical profile and cardiovascular baseline. Predictor fields spanning indices 9 through 15 were completely isolated from the amputation matrix (assigned a weight of 0).

- Linear Continuous Predictor Amputation (HDL Cholesterol): The missingness probability function was heavily conditioned on the primary target outcome alongside early continuous clinical vectors. Specifically, a strong probability anchor (weight of 0.4) was mapped to cardiac_outcome (outcome variable) to simulate clinical data collection omissions typically seen in active outcome cohorts. The remaining conditional probability weights were distributed uniformly (weight of 0.2) across predictors: age, sex, dm, htn, sbp, smoke, ckd, antiplatelet, ldl70),

- Non-Linear Continuous Predictor Imputation (eGFR): Similar to the linear continuum, missingness was anchored heavily to the primary event indicator, cardiac_outcome, with an identical weight of 0.4 to simulate clinical performance context trends. The remaining conditional dependencies were driven equally (weight of 0.2) across variables age, sex, dm, htn, sbp, dbp hdl, smoke, ckd, antiplatelet, ldl70

- Mixed Multivariate Scenarios: For mixed complexity amutations containing simultaneous variable omissions, the respective row-by-row structural regression matrices described above were executed concurrently. This exposed the model validation architecture to concurrent variations in categorical data streams and interdependent linear/non-linear continuous pathways.

**Supplementary Table S2. Amputation Patterns and Structural Predictor Configurations under Missing at Random (MAR) Scenarios.**

| Target Variable Subject to Amputation | Conditional Predictor Matrix | Assigned Amputation Matrix Regression Weight |
|---|---|---|
| Antiplatelet Use | cardiac_outcome*, age, sex, dm, htn, sbp, dbp, egfr | 0.2 (Uniformly Distributed) |
| | All remaining variables | 0.0 (Completely Isolated) |
| HDL Cholesterol | cardiac_outcome* | 0.4 (High Dependency Anchor) |
| | age, sex, dm, htn, sbp | 0.2 |
| | smoke, ckd, antiplatelet, ldl70 | 0.2 |
| | All remaining variables | 0.0 (Completely Isolated) |
| eGFR | cardiac_outcome* | 0.4 (High Dependency Anchor) |
| | age, sex, dm, htn, sbp, dbp | 0.2 |
| | hdl, smoke, ckd, antiplatelet, ldl70 | 0.2 |
| | All remaining variables | 0.0 (Completely Isolated) |

*outcome of interest

**Handling of Missing Data strategy**

1. **Complete Case Analysis (CCA):** As a reference standard for missing data mishandling, we performed a strict CCA by omitting any patient record with one or more missing values. To prevent perfect separation and zero-variance errors (singular Hessian matrices) during model fitting in smaller bootstrap samples, a dynamic safe-formula algorithm was applied to automatically exclude uninformative or perfectly separating variables prior to model fitting.
2. **Multiple Imputation by Chained Equations with Fully Conditional Specification (MICE-FCS):** Parametric multiple imputation was performed using standard FCS. Binary variables (e.g., sex, diabetes, hypertension) were imputed via logistic regression (logreg), and continuous variables (e.g., age, eGFR, lipids) via Bayesian linear regression (norm). We generated 10 imputed datasets ($m = 10$) with 5 iterations each. To handle the derived quadratic and interaction terms, we adopted the "Just Another Variable" (JAV) approach. The predictor matrix was manually modified to prevent collinearity loops (e.g., ensuring eGFR was not used to impute $eGFR^2$, and vice versa).
3. **Multiple Imputation with Predictive Mean Matching (MICE-PMM):** To account for potential non-normality in the clinical variables, a semi-parametric MICE approach was utilized. Predictive Mean Matching (pmm) was applied to both continuous and categorical variables, identifying the 10 closest donors for imputation. Similar to the FCS approach, 10 datasets were generated using 5 iterations, and the JAV approach was utilized for interaction and polynomial terms.

4. **Non-Parametric Iterative Random Forest (missForest):** A machine-learning imputation approach was implemented using the missForest algorithm, utilizing 50 trees (ntree = 50) and a maximum of 10 iterations. To rigorously prevent data leakage during model training, the primary outcome was deliberately excluded from the imputation predictor pool.
5. **k-Nearest Neighbors Imputation (kNN):** A distance-based single imputation method was applied using k = 10 neighbors. Missing continuous variables were imputed using the median of the neighbors, while missing categorical variables were imputed using the maximum category (mode). The outcome variable was also excluded from the distance calculations to prevent leakage.

**Supplementary Table S3.** Comparison of the standard deviation of imputed values across missingness scenarios and imputation methods.

| Missing | Scenario | Method | SD HDL Median (IQR) | SD eGFR Median (IQR) | SD eGFR_sq Median (IQR) | SD Age x eGFR |
|---|---|---|---|---|---|---|
| 30% Missing | Full dataset | CCA | 14.1 (13.5 , 14.7) | 24.6 (24.1 , 25.0) | 3276.2 (3209.5 , 3358.9) | 1509.9 (1477.0 , 1551.3) |
| | | KNN | 14.0 (13.5 , 14.7) | 25.3 (24.8 , 25.8) | 3409.0 (3316.3 , 3576.9) | 1567.8 (1523.0 , 1604.4) |
| | | FOREST | 14.0 (13.5 , 14.6) | 25.4 (24.9 , 25.8) | 3412.2 (3317.2 , 3565.2) | 1566.3 (1535.9 , 1604.1) |
| | | MICEpmm | 14.0 (13.4 , 14.5) | 25.3 (24.9 , 25.8) | 3410.5 (3317.2 , 3581.1) | 1566.6 (1539.4 , 1599.6) |
| | | MICEfcs | 14.1 (13.6 , 14.6) | 25.3 (24.7 , 25.9) | 3408.8 (3283.9 , 3554.8) | 1565.9 (1523.4 , 1602.9) |
| | LinCon | CCA | 14.1 (13.5 , 14.8) | 24.6 (23.9 , 25.2) | 3405.5 (3267.2 , 3595.9) | 1514.6 (1473.2 , 1556.2) |
| | | KNN | 12.5 (11.9 , 13.2) | 25.4 (24.8 , 25.8) | 3410.0 (3305.0 , 3613.0) | 1566.4 (1529.3 , 1596.4) |
| | | FOREST | 12.4 (11.9 , 13.0) | 25.3 (24.8 , 25.9) | 3418.1 (3305.1 , 3606.0) | 1562.8 (1529.5 , 1598.2) |
| | | MICEpmm | 14.1 (13.5 , 14.9) | 25.3 (24.7 , 25.9) | 3403.1 (3283.0 , 3531.6) | 1559.9 (1527.9 , 1603.2) |
| | | MICEfcs | 14.1 (13.4 , 14.9) | 25.3 (24.8 , 25.7) | 3415.9 (3292.4 , 3545.9) | 1562.1 (1525.1 , 1603.3) |
| | NLinCon | CCA | 13.3 (12.8 , 13.9) | 24.9 (24.2 , 25.4) | 3436.8 (3304.9 , 3601.3) | 1524.8 (1467.4 , 1562.1) |
| | | KNN | 14.1 (13.5 , 14.7) | 24.4 (23.7 , 25.9) | 3386.5 (3295.9 , 3568.9) | 1481.1 (1433.4 , 1555.3) |
| | | FOREST | 14.1 (13.5 , 14.5) | 21.3 (20.7 , 21.8) | 2948.2 (2836.7 , 3122.3) | 1312.2 (1267.7 , 1341.4) |

**Supplementary Table S3.** Comparison of the standard deviation of imputed values across missingness scenarios and imputation methods.

| Missing | Scenario | Method | SD HDL Median (IQR) | SD eGFR Median (IQR) | SD eGFR_sq Median (IQR) | SD Age x eGFR |
|---|---|---|---|---|---|---|
| | | MICEpmm | 14.1 (13.5 , 14.8) | 25.2 (24.4 , 25.8) | 3544.1 (3411.4 , 3732.6) | 1539.4 (1492.1 , 1584.0) |
| | | MICEfcs | 14.0 (13.5 , 14.7) | 25.0 (24.3 , 25.5) | 3475.2 (3332.5 , 3661.8) | 1515.3 (1469.3 , 1553.2) |
| | LinMix | CCA | 13.6 (13.1 , 14.1) | 24.7 (24.0 , 25.4) | 3419.8 (3256.6 , 3656.9) | 1515.8 (1474.2 , 1567.0) |
| | | KNN | 12.1 (11.6 , 12.6) | 25.4 (25.0 , 26.1) | 3427.8 (3319.9 , 3605.7) | 1567.1 (1530.3 , 1609.2) |
| | | FOREST | 11.9 (11.4 , 12.4) | 25.4 (25.0 , 25.9) | 3423.3 (3308.6 , 3591.9) | 1570.4 (1536.2 , 1608.4) |
| | | MICEpmm | 13.5 (13.0 , 14.2) | 25.3 (24.9 , 25.8) | 3410.5 (3307.6 , 3552.1) | 1562.6 (1534.6 , 1597.5) |
| | | MICEfcs | 13.7 (13.2 , 14.3) | 25.3 (24.7 , 25.8) | 3421.2 (3289.5 , 3580.0) | 1566.8 (1520.1 , 1605.5) |
| | NLinMix | CCA | 13.3 (12.9 , 13.8) | 24.6 (23.9 , 25.2) | 3413.1 (3253.0 , 3695.0) | 1509.8 (1473.7 , 1552.9) |
| | | KNN | 14.1 (13.6 , 14.8) | 24.7 (23.8 , 26.1) | 3426.6 (3291.3 , 3629.9) | 1483.2 (1434.3 , 1561.4) |
| | | FOREST | 14.1 (13.5 , 14.7) | 21.0 (20.1 , 21.7) | 2908.4 (2757.1 , 3158.9) | 1292.9 (1250.3 , 1347.2) |
| | | MICEpmm | 14.0 (13.5 , 14.7) | 25.0 (24.3 , 25.6) | 3576.5 (3400.9 , 3732.9) | 1526.5 (1475.6 , 1582.2) |
| | | MICEfcs | 14.1 (13.5 , 14.5) | 24.7 (24.0 , 25.4) | 3445.1 (3304.4 , 3659.0) | 1495.0 (1456.7 , 1548.7) |
| | All Conditions | CCA | 13.6 (13.2 , 14.2) | 24.9 (24.3 , 25.6) | 3411.2 (3278.3 , 3625.4) | 1533.4 (1487.2 , 1589.6) |
| | | KNN | 12.4 (11.9 , 13.2) | 24.8 (24.1 , 26.6) | 3408.3 (3268.8 , 3612.1) | 1489.2 (1449.5 , 1552.4) |

**Supplementary Table S3.** Comparison of the standard deviation of imputed values across missingness scenarios and imputation methods.

| Missing | Scenario | Method | SD HDL Median (IQR) | SD eGFR Median (IQR) | SD eGFR_sq Median (IQR) | SD Age x eGFR |
|---|---|---|---|---|---|---|
| | | FOREST | 11.9 (11.4 , 12.4) | 21.3 (20.8 , 22.1) | 2936.7 (2809.3 , 3129.7) | 1322.8 (1285.5 , 1366.9) |
| | | MICEpmm | 13.5 (13.0 , 14.0) | 25.4 (24.7 , 26.0) | 3585.3 (3415.7 , 3770.7) | 1558.4 (1507.5 , 1602.6) |
| | | MICEfcs | 14.0 (13.4 , 14.6) | 25.1 (24.4 , 25.7) | 3480.0 (3307.0 , 3668.7) | 1527.3 (1492.3 , 1566.1) |
| 60% Missing | Full dataset | CCA | 14.4 (13.7 , 15.6) | 24.8 (24.0 , 25.5) | 3348.9 (3251.8 , 3452.7) | 1516.2 (1455.6 , 1563.8) |
| | | KNN | 14.1 (13.5 , 14.6) | 25.4 (24.8 , 26.0) | 3426.0 (3299.0 , 3548.2) | 1569.3 (1520.6 , 1602.3) |
| | | FOREST | 14.0 (13.5 , 14.5) | 25.3 (24.9 , 25.8) | 3408.3 (3316.4 , 3551.5) | 1562.9 (1533.1 , 1590.8) |
| | | MICEpmm | 14.1 (13.6 , 14.6) | 25.3 (24.9 , 25.8) | 3420.3 (3308.3 , 3566.8) | 1562.4 (1532.9 , 1601.2) |
| | | MICEfcs | 14.0 (13.5 , 14.5) | 25.4 (24.7 , 25.9) | 3422.8 (3296.4 , 3572.8) | 1566.6 (1531.2 , 1601.1) |
| | LinCon | CCA | 14.1 (13.3 , 14.9) | 23.8 (23.0 , 24.4) | 3326.2 (3232.6 , 3434.1) | 1442.3 (1383.8 , 1485.6) |
| | | KNN | 10.8 (10.2 , 11.3) | 25.4 (24.7 , 26.0) | 3424.9 (3293.9 , 3624.8) | 1569.4 (1528.2 , 1606.7) |
| | | FOREST | 10.4 (9.8 , 10.9) | 25.3 (24.9 , 25.9) | 3408.2 (3316.8 , 3583.4) | 1566.3 (1532.2 , 1602.3) |
| | | MICEpmm | 14.0 (13.2 , 14.8) | 25.3 (24.8 , 25.8) | 3405.5 (3307.6 , 3574.2) | 1566.5 (1527.0 , 1598.5) |
| | | MICEfcs | 14.4 (13.3 , 15.1) | 25.4 (24.8 , 25.8) | 3424.8 (3289.0 , 3608.4) | 1566.4 (1529.2 , 1605.4) |
| | NLinCon | CCA | 13.5 (12.9 , 14.2) | 24.1 (23.3 , 25.1) | 3323.5 (3203.7 , 3476.3) | 1472.7 (1424.1 , 1530.4) |

**Supplementary Table S3.** Comparison of the standard deviation of imputed values across missingness scenarios and imputation methods.

| Missing | Scenario | Method | SD HDL Median (IQR) | SD eGFR Median (IQR) | SD eGFR_sq Median (IQR) | SD Age x eGFR |
|---|---|---|---|---|---|---|
| | | KNN | 14.0 (13.6 , 14.5) | 22.3 (21.3 , 23.3) | 3153.4 (3019.1 , 3419.9) | 1368.1 (1301.0 , 1449.2) |
| | | FOREST | 14.0 (13.6 , 14.6) | 16.8 (15.6 , 18.1) | 2317.0 (2141.1 , 2495.7) | 1044.9 (975.3 , 1117.7) |
| | | MICEpmm | 14.0 (13.6 , 14.6) | 24.7 (23.8 , 25.8) | 3557.1 (3363.7 , 3738.7) | 1487.6 (1442.6 , 1573.2) |
| | | MICEfcs | 14.1 (13.5 , 14.6) | 24.7 (23.9 , 25.5) | 3448.8 (3316.4 , 3610.4) | 1470.7 (1413.4 , 1516.3) |
| | LinMix | CCA | 13.0 (12.4 , 13.9) | 23.7 (22.9 , 24.7) | 3282.2 (3134.7 , 3460.5) | 1445.3 (1391.8 , 1499.3) |
| | | KNN | 10.1 (9.5 , 10.9) | 25.4 (24.8 , 25.9) | 3408.2 (3294.0 , 3600.3) | 1566.1 (1534.1 , 1607.9) |
| | | FOREST | 9.4 (9.0 , 10.0) | 25.4 (24.8 , 26.0) | 3411.7 (3301.9 , 3602.1) | 1568.4 (1530.2 , 1608.6) |
| | | MICEpmm | 13.0 (12.3 , 14.1) | 25.3 (24.8 , 25.8) | 3400.1 (3310.2 , 3567.6) | 1561.9 (1527.4 , 1602.7) |
| | | MICEfcs | 13.4 (12.7 , 14.5) | 25.3 (24.8 , 26.0) | 3411.5 (3297.0 , 3587.7) | 1569.6 (1521.0 , 1603.5) |
| | NLinMix | CCA | 13.0 (12.5 , 13.5) | 23.5 (23.0 , 24.4) | 3293.1 (3200.3 , 3424.6) | 1436.9 (1394.4 , 1486.5) |
| | | KNN | 14.1 (13.4 , 14.5) | 21.4 (20.5 , 22.5) | 3410.6 (3299.9 , 3584.5) | 1565.6 (1531.6 , 1611.9) |
| | | FOREST | 14.0 (13.5 , 14.6) | 23.9 (23.4 , 24.6) | 3412.5 (3300.4 , 3557.9) | 1563.3 (1533.4 , 1600.0) |
| | | MICEpmm | 14.0 (13.4 , 14.6) | 25.6 (25.0 , 26.2) | 3401.6 (3309.9 , 3605.2) | 1564.9 (1535.1 , 1605.1) |
| | | MICEfcs | 14.0 (13.5 , 14.6) | 25.9 (25.3 , 26.4) | 3396.7 (3299.7 , 3572.6) | 1565.3 (1534.8 , 1598.2) |

**Supplementary Table S3.** Comparison of the standard deviation of imputed values across missingness scenarios and imputation methods.

| Missing | Scenario | Method | SD HDL Median (IQR) | SD eGFR Median (IQR) | SD eGFR_sq Median (IQR) | SD Age x eGFR |
|---|---|---|---|---|---|---|
| | All Conditions | CCA | 13.3 (12.7 , 13.9) | 24.8 (24.1 , 25.6) | 3364.4 (3268.2 , 3490.7) | 1516.3 (1466.7 , 1561.4) |
| | | KNN | 10.4 (9.7 , 11.2) | 22.6 (21.4 , 24.2) | 3193.6 (3018.7 , 3421.5) | 1381.7 (1298.8 , 1481.5) |
| | | FOREST | 9.5 (9.1 , 10.0) | 17.5 (16.2 , 18.7) | 2390.5 (2193.9 , 2557.4) | 1075.3 (1002.0 , 1152.0) |
| | | MICEpmm | 13.2 (12.5 , 13.8) | 25.4 (24.6 , 26.6) | 3641.8 (3486.5 , 3835.1) | 1531.3 (1474.5 , 1628.4) |
| | | MICEfcs | 13.6 (12.8 , 14.2) | 25.2 (24.3 , 26.0) | 3476.9 (3363.9 , 3619.1) | 1512.8 (1445.9 , 1570.2) |
| 90% Missing | Full dataset | CCA | 13.7 (12.3 , 15.9) | 24.2 (22.4 , 25.5) | 3276.4 (3049.4 , 3455.1) | 1460.8 (1354.7 , 1552.7) |
| | | KNN | 14.1 (13.7 , 14.6) | 25.3 (24.9 , 25.9) | 3416.7 (3317.5 , 3555.1) | 1566.6 (1536.7 , 1607.4) |
| | | FOREST | 14.0 (13.6 , 14.5) | 25.3 (24.9 , 25.9) | 3409.4 (3319.1 , 3575.9) | 1564.6 (1540.0 , 1598.0) |
| | | MICEpmm | 14.0 (13.6 , 14.6) | 25.4 (24.9 , 26.0) | 3414.8 (3307.6 , 3584.9) | 1569.0 (1534.5 , 1606.4) |
| | | MICEfcs | 14.0 (13.5 , 14.6) | 25.3 (24.8 , 25.9) | 3414.3 (3291.2 , 3543.4) | 1567.8 (1529.9 , 1600.7) |
| | LinCon | CCA | 14.5 (12.9 , 16.1) | 23.1 (22.0 , 24.2) | 3292.7 (3129.5 , 3433.0) | 1349.2 (1263.4 , 1427.5) |
| | | KNN | 8.7 (7.6 , 9.9) | 25.3 (24.8 , 25.9) | 3422.1 (3288.3 , 3625.4) | 1561.6 (1528.1 , 1609.4) |
| | | FOREST | 7.9 (6.9 , 8.9) | 25.3 (24.8 , 25.7) | 3407.3 (3305.3 , 3567.2) | 1564.8 (1526.6 , 1593.5) |
| | | MICEpmm | 14.6 (13.0 , 16.5) | 25.3 (24.7 , 25.8) | 3426.0 (3289.9 , 3570.8) | 1565.1 (1526.7 , 1606.2) |

**Supplementary Table S3.** Comparison of the standard deviation of imputed values across missingness scenarios and imputation methods.

| Missing | Scenario | Method | SD HDL Median (IQR) | SD eGFR Median (IQR) | SD eGFR_sq Median (IQR) | SD Age x eGFR |
|---|---|---|---|---|---|---|
| | | MICEfcs | 15.6 (13.5 , 18.2) | 25.4 (24.9 , 25.8) | 3420.4 (3318.2 , 3574.5) | 1569.3 (1533.5 , 1598.3) |
| | NLinCon | CCA | 12.8 (11.7 , 13.8) | 25.1 (23.7 , 26.6) | 3471.5 (3205.9 , 3672.9) | 1486.7 (1390.9 , 1581.6) |
| | | KNN | 14.1 (13.4 , 14.5) | 14.8 (12.5 , 18.8) | 2244.7 (1927.9 , 3097.0) | 930.9 (799.9 , 1098.7) |
| | | FOREST | 14.1 (13.7 , 14.7) | 11.6 (9.7 , 16.8) | 1581.5 (1270.7 , 2283.9) | 689.8 (587.6 , 968.8) |
| | | MICEpmm | 14.0 (13.5 , 14.6) | 23.9 (21.9 , 25.6) | 3364.2 (3119.7 , 3688.0) | 1374.8 (1233.2 , 1490.9) |
| | | MICEfcs | 14.1 (13.5 , 14.8) | 25.8 (24.0 , 27.4) | 3689.7 (3316.1 , 4030.9) | 1506.6 (1386.1 , 1615.4) |
| | LinMix | CCA | 14.2 (12.9 , 15.9) | 23.3 (21.6 , 24.9) | 3234.5 (3020.4 , 3450.7) | 1462.2 (1348.2 , 1556.2) |
| | | KNN | 8.4 (7.1 , 10.3) | 25.4 (24.7 , 25.9) | 3416.3 (3291.7 , 3572.5) | 1569.1 (1520.9 , 1605.1) |
| | | FOREST | 7.3 (6.5 , 8.5) | 25.3 (24.8 , 25.8) | 3415.3 (3299.4 , 3588.3) | 1565.8 (1529.1 , 1594.3) |
| | | MICEpmm | 13.7 (12.5 , 15.8) | 25.3 (24.8 , 25.9) | 3404.2 (3282.2 , 3575.9) | 1563.0 (1524.6 , 1606.3) |
| | | MICEfcs | 15.3 (13.2 , 17.1) | 25.3 (24.8 , 25.8) | 3417.9 (3310.1 , 3580.7) | 1567.5 (1523.4 , 1600.6) |
| | NLinMix | CCA | 12.7 (11.6 , 13.5) | 23.5 (22.4 , 24.8) | 3282.8 (3122.2 , 3534.7) | 1435.4 (1341.5 , 1529.5) |
| | | KNN | 14.1 (13.6 , 14.6) | 15.6 (12.8 , 18.3) | 2410.8 (2039.3 , 2915.2) | 990.3 (827.6 , 1197.6) |
| | | FOREST | 14.0 (13.7 , 14.6) | 14.1 (8.7 , 18.0) | 1805.0 (1187.6 , 2478.5) | 787.2 (538.4 , 1042.5) |

**Supplementary Table S3.** Comparison of the standard deviation of imputed values across missingness scenarios and imputation methods.

| Missing | Scenario | Method | SD HDL Median (IQR) | SD eGFR Median (IQR) | SD eGFR_sq Median (IQR) | SD Age x eGFR |
|---|---|---|---|---|---|---|
| | | MICEpmm | 14.0 (13.5 , 14.6) | 22.1 (20.7 , 23.7) | 3257.3 (3011.6 , 3550.7) | 1278.0 (1187.2 , 1409.8) |
| | | MICEfcs | 14.0 (13.6 , 14.7) | 24.5 (23.1 , 26.2) | 3498.7 (3277.7 , 3771.9) | 1455.0 (1349.3 , 1578.8) |
| | All Conditions | CCA | 13.0 (12.0 , 13.9) | 23.9 (22.5 , 25.5) | 3219.9 (3048.9 , 3374.0) | 1477.8 (1387.0 , 1577.3) |
| | | KNN | 9.1 (7.4 , 10.8) | 13.1 (11.5 , 15.4) | 2175.0 (1855.3 , 2658.9) | 845.0 (709.9 , 1049.9) |
| | | FOREST | 6.5 (5.8 , 7.3) | 11.0 (8.3 , 15.2) | 1488.7 (1107.2 , 2183.5) | 678.2 (549.1 , 995.8) |
| | | MICEpmm | 13.4 (12.0 , 14.6) | 23.3 (21.4 , 25.7) | 3208.1 (2976.6 , 3483.6) | 1379.6 (1259.8 , 1527.3) |
| | | MICEfcs | 14.3 (12.5 , 16.3) | 25.3 (23.5 , 27.2) | 3466.2 (3226.2 , 3774.3) | 1465.8 (1361.9 , 1597.4) |

**Supplementary Table S4.** Comparison of the median of imputed values across missingness scenarios and imputation methods.

| Missing | Scenario | Method | HDL Median (IQR) | eGFR Median (IQR) | eGFR_sq Median (IQR) | Age x eGFR Median (IQR) |
|---|---|---|---|---|---|---|
| 30% Missing | Full dataset | CCA | 48.0 (47.0 , 49.0) | 70.2 (69.6 , 71.6) | 4929.8 (4841.8 , 5121.8) | 4528.0 (4451.1 , 4613.6) |
| | | KNN | 47.0 (47.0 , 48.0) | 69.6 (68.3 , 70.1) | 4841.8 (4660.4 , 4911.2) | 4477.1 (4409.8 , 4554.8) |
| | | FOREST | 47.0 (47.0 , 48.0) | 69.2 (68.3 , 70.1) | 4784.5 (4660.4 , 4910.3) | 4477.1 (4403.4 , 4547.5) |
| | | MICEpmm | 47.0 (47.0 , 48.0) | 69.3 (68.3 , 70.4) | 4806.5 (4660.4 , 4958.0) | 4477.1 (4411.7 , 4549.4) |
| | | MICEfcs | 47.0 (47.0 , 48.0) | 69.4 (68.5 , 70.1) | 4811.5 (4687.1 , 4910.3) | 4477.1 (4415.3 , 4549.4) |
| | LinCon | CCA | 48.0 (47.0 , 48.0) | 71.0 (70.1 , 72.6) | 5034.4 (4910.3 , 5267.7) | 4589.4 (4521.8 , 4642.8) |
| | | KNN | 47.5 (47.0 , 48.0) | 69.3 (68.3 , 70.1) | 4801.4 (4668.8 , 4912.0) | 4477.1 (4411.7 , 4549.4) |
| | | FOREST | 48.0 (47.7 , 48.7) | 69.2 (68.4 , 70.1) | 4793.2 (4678.2 , 4910.3) | 4477.1 (4407.3 , 4541.1) |
| | | MICEpmm | 47.0 (47.0 , 48.0) | 69.2 (68.5 , 70.1) | 4793.2 (4687.1 , 4910.3) | 4477.1 (4411.7 , 4549.4) |
| | | MICEfcs | 48.0 (47.2 , 48.6) | 69.1 (68.4 , 70.1) | 4775.9 (4678.2 , 4910.3) | 4475.6 (4411.7 , 4530.0) |
| | NLinCon | CCA | 47.0 (47.0 , 47.5) | 71.1 (69.8 , 72.3) | 5050.3 (4869.8 , 5223.7) | 4563.8 (4477.1 , 4620.5) |
| | | KNN | 47.0 (47.0 , 48.0) | 76.5 (74.7 , 79.5) | 5683.2 (5494.4 , 6062.3) | 4780.8 (4686.3 , 4942.1) |
| | | FOREST | 47.0 (47.0 , 48.0) | 70.4 (69.2 , 71.6) | 4948.6 (4792.0 , 5110.7) | 4551.1 (4445.9 , 4633.0) |
| | | MICEpmm | 47.0 (47.0 , 48.0) | 68.6 (67.7 , 69.7) | 4774.3 (4643.6 , 4910.3) | 4424.7 (4343.9 , 4487.0) |
| | | MICEfcs | 47.0 (47.0 , 48.0) | 70.1 (69.1 , 71.3) | 5092.5 (4948.6 , 5308.7) | 4534.0 (4457.2 , 4629.4) |

**Supplementary Table S4.** Comparison of the median of imputed values across missingness scenarios and imputation methods.

| Missing | Scenario | Method | HDL Median (IQR) | eGFR Median (IQR) | eGFR_sq Median (IQR) | Age x eGFR Median (IQR) |
|---|---|---|---|---|---|---|
| | LinMix | CCA | 47.0 (47.0 , 47.5) | 71.1 (70.1 , 72.4) | 5050.3 (4910.3 , 5234.7) | 4577.5 (4504.8 , 4642.6) |
| | | KNN | 47.0 (46.0 , 47.0) | 69.3 (68.3 , 70.1) | 4797.3 (4664.2 , 4910.3) | 4477.1 (4389.5 , 4534.5) |
| | | FOREST | 47.0 (46.7 , 48.0) | 69.3 (68.3 , 70.1) | 4797.3 (4660.4 , 4910.3) | 4477.1 (4416.5 , 4549.4) |
| | | MICEpmm | 47.0 (46.0 , 47.0) | 69.4 (68.3 , 70.1) | 4811.5 (4668.8 , 4910.3) | 4477.1 (4411.7 , 4549.4) |
| | | MICEfcs | 47.0 (47.0 , 48.0) | 69.4 (68.3 , 70.1) | 4811.5 (4660.4 , 4912.0) | 4479.2 (4413.9 , 4554.8) |
| | NLinMix | CCA | 47.0 (46.0 , 47.8) | 71.1 (70.1 , 72.6) | 5050.3 (4910.3 , 5273.1) | 4582.1 (4502.7 , 4635.2) |
| | | KNN | 47.0 (47.0 , 48.0) | 78.6 (76.0 , 82.7) | 5977.7 (5603.9 , 6462.6) | 4897.0 (4742.1 , 5069.3) |
| | | FOREST | 47.0 (47.0 , 48.0) | 70.5 (69.0 , 71.8) | 4955.1 (4795.3 , 5158.9) | 4540.5 (4430.6 , 4661.1) |
| | | MICEpmm | 47.0 (47.0 , 48.0) | 68.3 (67.3 , 69.2) | 4744.4 (4595.3 , 4861.0) | 4423.3 (4342.3 , 4516.0) |
| | | MICEfcs | 47.0 (47.0 , 48.0) | 70.1 (69.1 , 71.1) | 5129.3 (4974.3 , 5318.1) | 4549.6 (4473.4 , 4627.5) |
| | All Conditions | CCA | 47.0 (47.0 , 48.0) | 70.1 (69.2 , 71.5) | 4912.0 (4795.0 , 5114.1) | 4519.0 (4449.0 , 4590.7) |
| | | KNN | 48.0 (47.0 , 49.0) | 77.5 (75.6 , 80.4) | 5850.6 (5509.3 , 6274.2) | 4846.6 (4741.1 , 4985.4) |
| | | FOREST | 47.0 (46.9 , 48.0) | 70.1 (68.6 , 71.2) | 4879.6 (4692.2 , 5050.6) | 4470.9 (4402.3 , 4577.2) |
| | | MICEpmm | 47.0 (46.0 , 47.0) | 67.7 (66.8 , 69.1) | 4660.4 (4514.1 , 4827.4) | 4381.7 (4323.6 , 4471.5) |
| | | MICEfcs | 47.0 (47.0 , 48.0) | 69.8 (68.6 , 70.8) | 5073.2 (4916.1 , 5246.8) | 4516.0 (4424.7 , 4592.2) |
| 60% Missing | Full dataset | CCA | 48.0 (47.0 , 48.0) | 71.6 (70.3 , 73.1) | 5122.2 (4938.5 , 5342.2) | 4533.0 (4437.1 , 4625.0) |

**Supplementary Table S4.** Comparison of the median of imputed values across missingness scenarios and imputation methods.

| Missing | Scenario | Method | HDL Median (IQR) | eGFR Median (IQR) | eGFR_sq Median (IQR) | Age x eGFR Median (IQR) |
|---|---|---|---|---|---|---|
| | | KNN | 47.0 (47.0 , 48.0) | 69.3 (68.4 , 70.1) | 4801.4 (4678.2 , 4910.3) | 4477.1 (4411.7 , 4545.4) |
| | | FOREST | 47.0 (47.0 , 48.0) | 69.2 (68.5 , 70.1) | 4793.2 (4687.8 , 4910.3) | 4477.1 (4413.1 , 4545.1) |
| | | MICEpmm | 47.0 (47.0 , 48.0) | 69.2 (68.3 , 70.4) | 4793.2 (4660.4 , 4956.0) | 4477.1 (4411.7 , 4567.3) |
| | | MICEfcs | 47.0 (47.0 , 48.0) | 69.5 (68.4 , 70.5) | 4826.7 (4682.8 , 4965.0) | 4481.9 (4417.8 , 4579.5) |
| | LinCon | CCA | 48.0 (47.0 , 49.0) | 73.8 (72.8 , 75.7) | 5448.6 (5292.8 , 5731.0) | 4674.0 (4592.5 , 4791.4) |
| | | KNN | 47.5 (47.0 , 48.0) | 69.6 (68.6 , 70.2) | 4841.8 (4709.3 , 4930.4) | 4481.9 (4403.6 , 4549.4) |
| | | FOREST | 48.4 (47.8 , 49.1) | 69.4 (68.3 , 70.1) | 4811.5 (4660.4 , 4910.3) | 4477.1 (4407.3 , 4535.2) |
| | | MICEpmm | 47.0 (46.5 , 49.0) | 69.1 (68.4 , 70.1) | 4775.9 (4682.8 , 4910.3) | 4477.1 (4410.7 , 4554.8) |
| | | MICEfcs | 48.9 (48.0 , 49.5) | 69.5 (68.3 , 70.1) | 4826.7 (4660.4 , 4910.3) | 4481.9 (4424.7 , 4549.4) |
| | NLinCon | CCA | 46.0 (46.0 , 47.0) | 72.6 (71.1 , 73.4) | 5267.7 (5050.3 , 5385.3) | 4592.5 (4487.7 , 4681.9) |
| | | KNN | 47.0 (47.0 , 48.0) | 88.1 (85.2 , 89.9) | 7208.6 (6648.0 , 7566.8) | 5243.6 (5087.3 , 5358.0) |
| | | FOREST | 47.0 (47.0 , 48.0) | 72.0 (70.3 , 73.4) | 5163.0 (4935.1 , 5397.7) | 4528.0 (4435.2 , 4720.3) |
| | | MICEpmm | 47.0 (47.0 , 48.0) | 67.3 (65.8 , 69.2) | 4660.4 (4459.6 , 4910.3) | 4323.6 (4256.9 , 4424.7) |
| | | MICEfcs | 47.0 (47.0 , 48.0) | 70.1 (68.9 , 71.7) | 5308.7 (5103.0 , 5525.7) | 4549.4 (4450.8 , 4643.3) |
| | LinMix | CCA | 47.0 (46.0 , 47.0) | 71.8 (70.3 , 73.7) | 5161.6 (4948.4 , 5429.0) | 4585.9 (4484.8 , 4665.3) |
| | | KNN | 47.0 (46.0 , 48.0) | 69.4 (68.3 , 70.1) | 4811.5 (4660.4 , 4911.2) | 4477.1 (4411.7 , 4528.9) |

**Supplementary Table S4.** Comparison of the median of imputed values across missingness scenarios and imputation methods.

| Missing | Scenario | Method | HDL Median (IQR) | eGFR Median (IQR) | eGFR_sq Median (IQR) | Age x eGFR Median (IQR) |
|---|---|---|---|---|---|---|
| | | FOREST | 46.9 (46.1 , 47.3) | 69.4 (68.3 , 70.2) | 4811.5 (4660.4 , 4930.4) | 4480.6 (4407.3 , 4554.8) |
| | | MICEpmm | 46.0 (46.0 , 47.0) | 69.4 (68.3 , 70.1) | 4811.5 (4668.8 , 4910.3) | 4479.3 (4421.6 , 4554.8) |
| | | MICEfcs | 47.0 (46.1 , 48.0) | 69.4 (68.4 , 70.3) | 4811.5 (4678.2 , 4938.3) | 4478.1 (4411.7 , 4554.8) |
| | NLinMix | CCA | 47.0 (46.2 , 48.0) | 72.4 (70.8 , 74.0) | 5234.7 (5018.6 , 5481.1) | 4592.5 (4527.5 , 4687.7) |
| | | KNN | 47.0 (47.0 , 48.0) | 69.1 (68.0 , 70.7) | 4793.2 (4668.8 , 4911.2) | 4476.5 (4411.7 , 4552.2) |
| | | FOREST | 47.0 (47.0 , 48.0) | 69.2 (68.1 , 70.1) | 4811.5 (4682.8 , 4930.4) | 4479.2 (4411.7 , 4554.8) |
| | | MICEpmm | 47.0 (47.0 , 48.0) | 69.6 (68.1 , 70.1) | 4826.7 (4660.4 , 4910.3) | 4477.1 (4402.5 , 4549.4) |
| | | MICEfcs | 47.0 (47.0 , 48.0) | 69.7 (68.6 , 70.7) | 4801.4 (4651.6 , 4910.3) | 4477.1 (4407.3 , 4545.1) |
| | All Conditions | CCA | 47.0 (46.0 , 48.0) | 70.6 (69.0 , 71.8) | 4981.5 (4766.8 , 5161.6) | 4516.3 (4417.8 , 4599.6) |
| | | KNN | 49.0 (47.0 , 51.0) | 89.9 (88.3 , 91.8) | 7755.7 (7332.2 , 8084.0) | 5410.1 (5276.2 , 5536.0) |
| | | FOREST | 47.9 (47.0 , 48.6) | 70.3 (68.0 , 72.2) | 4888.2 (4639.8 , 5184.6) | 4428.7 (4367.0 , 4562.3) |
| | | MICEpmm | 47.0 (46.0 , 48.0) | 66.3 (64.9 , 67.7) | 4514.9 (4327.1 , 4708.5) | 4274.7 (4158.9 , 4404.4) |
| | | MICEfcs | 48.0 (47.0 , 48.9) | 69.1 (67.5 , 70.9) | 5161.6 (4928.0 , 5448.6) | 4474.6 (4385.1 , 4609.2) |
| 90% Missing | Full dataset | CCA | 49.0 (48.0 , 50.0) | 70.3 (68.3 , 73.0) | 4948.6 (4660.4 , 5326.3) | 4369.1 (4300.8 , 4592.5) |
| | | KNN | 47.0 (47.0 , 48.0) | 69.5 (68.3 , 70.1) | 4826.7 (4668.8 , 4910.3) | 4481.9 (4413.1 , 4552.2) |
| | | FOREST | 47.0 (47.0 , 48.0) | 69.3 (68.3 , 70.1) | 4801.4 (4660.4 , 4910.3) | 4477.1 (4415.3 , 4552.2) |

**Supplementary Table S4.** Comparison of the median of imputed values across missingness scenarios and imputation methods.

| Missing | Scenario | Method | HDL Median (IQR) | eGFR Median (IQR) | eGFR_sq Median (IQR) | Age x eGFR Median (IQR) |
|---|---|---|---|---|---|---|
| | | MICEpmm | 47.0 (47.0 , 48.0) | 69.2 (68.3 , 70.1) | 4784.5 (4660.4 , 4912.0) | 4475.8 (4409.8 , 4554.8) |
| | | MICEfcs | 47.0 (47.0 , 48.0) | 69.4 (68.4 , 70.1) | 4811.5 (4682.8 , 4912.0) | 4477.1 (4411.7 , 4552.2) |
| | LinCon | CCA | 48.5 (47.0 , 50.0) | 74.9 (72.4 , 77.4) | 5603.9 (5244.6 , 5984.3) | 4639.5 (4534.0 , 4759.7) |
| | | KNN | 47.5 (46.0 , 50.0) | 69.4 (68.1 , 70.1) | 4811.5 (4643.6 , 4911.2) | 4477.1 (4410.4 , 4549.4) |
| | | FOREST | 49.0 (48.0 , 50.3) | 69.4 (68.3 , 70.1) | 4811.5 (4660.4 , 4912.0) | 4477.1 (4417.8 , 4543.1) |
| | | MICEpmm | 48.0 (45.0 , 50.0) | 69.4 (68.6 , 70.3) | 4811.5 (4707.7 , 4947.6) | 4479.2 (4403.6 , 4554.8) |
| | | MICEfcs | 48.7 (46.8 , 50.7) | 69.4 (68.3 , 70.1) | 4811.5 (4668.8 , 4910.3) | 4481.9 (4403.4 , 4557.9) |
| | NLinCon | CCA | 46.0 (44.2 , 47.0) | 72.9 (70.0 , 76.0) | 5319.6 (4897.4 , 5774.7) | 4559.6 (4406.7 , 4757.0) |
| | | KNN | 47.0 (47.0 , 48.0) | 89.2 (85.0 , 95.0) | 7209.7 (6626.3 , 8093.8) | 5088.4 (4866.9 , 5370.4) |
| | | FOREST | 47.0 (47.0 , 48.0) | 73.5 (70.5 , 76.6) | 5148.4 (4820.2 , 5738.7) | 4666.2 (4528.6 , 4861.9) |
| | | MICEpmm | 47.0 (47.0 , 48.0) | 64.9 (61.9 , 69.0) | 4514.9 (3981.9 , 5110.8) | 4231.7 (4062.2 , 4489.2) |
| | | MICEfcs | 47.0 (47.0 , 48.0) | 69.8 (67.4 , 71.9) | 5465.6 (5074.1 , 5794.9) | 4510.8 (4340.3 , 4649.4) |
| | LinMix | CCA | 45.0 (44.0 , 47.0) | 72.7 (69.9 , 74.9) | 5288.5 (4889.9 , 5609.7) | 4591.0 (4386.5 , 4771.7) |
| | | KNN | 46.0 (44.0 , 47.5) | 69.4 (68.1 , 70.1) | 4811.5 (4643.6 , 4911.2) | 4481.9 (4396.3 , 4549.4) |
| | | FOREST | 46.3 (45.0 , 47.6) | 69.2 (68.3 , 70.1) | 4784.5 (4660.4 , 4910.3) | 4475.8 (4403.4 , 4529.1) |
| | | MICEpmm | 45.0 (43.0 , 46.0) | 69.4 (68.5 , 70.1) | 4811.5 (4692.2 , 4910.3) | 4478.2 (4411.7 , 4545.1) |

**Supplementary Table S4.** Comparison of the median of imputed values across missingness scenarios and imputation methods.

| Missing | Scenario | Method | HDL Median (IQR) | eGFR Median (IQR) | eGFR_sq Median (IQR) | Age x eGFR Median (IQR) |
|---|---|---|---|---|---|---|
| | | MICEfcs | 46.3 (44.0 , 48.6) | 69.4 (68.4 , 70.1) | 4811.5 (4673.0 , 4910.3) | 4479.2 (4402.7 , 4552.2) |
| | NLinMix | CCA | 45.0 (44.0 , 46.0) | 73.6 (70.2 , 77.5) | 5417.8 (4926.6 , 6000.1) | 4560.5 (4396.3 , 4701.0) |
| | | KNN | 47.0 (47.0 , 48.0) | 94.5 (90.2 , 97.2) | 8170.4 (7309.3 , 9025.9) | 5349.3 (5018.9 , 5713.1) |
| | | FOREST | 47.0 (47.0 , 48.0) | 73.2 (66.2 , 78.1) | 5258.8 (4441.5 , 6067.0) | 4567.6 (4322.7 , 4923.3) |
| | | MICEpmm | 47.0 (47.0 , 48.0) | 65.3 (63.3 , 68.9) | 4513.4 (4207.2 , 4999.5) | 4262.7 (4116.5 , 4431.2) |
| | | MICEfcs | 47.0 (47.0 , 48.0) | 70.2 (67.4 , 73.1) | 5536.4 (5180.6 , 6037.7) | 4523.5 (4332.2 , 4668.1) |
| | All Conditions | CCA | 46.0 (45.0 , 48.0) | 72.1 (70.2 , 76.3) | 5194.2 (4931.2 , 5817.9) | 4650.3 (4512.3 , 4795.0) |
| | | KNN | 47.0 (43.5 , 51.0) | 91.1 (87.6 , 95.6) | 7633.5 (6687.8 , 8852.6) | 5345.0 (5018.8 , 5705.1) |
| | | FOREST | 48.0 (46.3 , 49.8) | 74.0 (70.0 , 78.3) | 5306.0 (4902.5 , 6014.4) | 4799.3 (4523.6 , 5037.9) |
| | | MICEpmm | 46.0 (44.0 , 48.0) | 66.2 (62.7 , 70.1) | 4513.4 (4075.9 , 5018.6) | 4341.8 (4187.1 , 4535.2) |
| | | MICEfcs | 47.8 (45.2 , 50.6) | 70.7 (67.9 , 73.0) | 5508.9 (5062.3 , 5856.9) | 4629.7 (4473.3 , 4790.6) |

**Supplementary Table S5.** Number of retained predictors in model and distribution of Linear predictor (LP)

| Missing | Scenario | Method | Retained Variables<br>Median (IQR) | Mean LP<br>Median (IQR) | SD LP<br>Median (IQR) |
|---|---|---|---|---|---|
| 30% Missing | Categorical | CCA | 8.0<br>(6.0 , 10.5) | -2.72<br>(-2.89 , -2.54) | 0.96<br>(0.81 , 1.11) |
| | | KNN | 9.0<br>(6.0 , 11.0) | -2.17<br>(-2.30 , -2.06) | 0.93<br>(0.85 , 1.03) |
| | | FOREST | 9.0<br>(7.0 , 11.0) | -2.20<br>(-2.31 , -2.09) | 0.96<br>(0.83 , 1.06) |
| | | MICEpmm | 10.0<br>(8.0 , 12.0) | -2.19<br>(-2.32 , -2.05) | 0.93<br>(0.84 , 1.04) |
| | | MICEfcs | 10.0<br>(8.0 , 12.0) | -2.19<br>(-2.29 , -2.09) | 0.93<br>(0.86 , 1.03) |
| | LinCon | CCA | 7.0<br>(3.5 , 10.0) | -2.80<br>(-2.99 , -2.65) | 1.00<br>(0.86 , 1.17) |
| | | KNN | 9.0<br>(7.0 , 12.0) | -2.20<br>(-2.31 , -2.10) | 0.97<br>(0.89 , 1.06) |
| | | FOREST | 8.0<br>(7.0 , 10.0) | -2.20<br>(-2.29 , -2.10) | 0.96<br>(0.87 , 1.05) |
| | | MICEpmm | 10.0<br>(7.0 , 13.0) | -2.21<br>(-2.33 , -2.11) | 0.97<br>(0.88 , 1.08) |
| | | MICEfcs | 10.0<br>(8.0 , 13.0) | -2.21<br>(-2.31 , -2.09) | 0.98<br>(0.88 , 1.09) |
| | NLinCon | CCA | 9.0<br>(5.0 , 11.0) | -2.71<br>(-2.87 , -2.56) | 0.89<br>(0.75 , 1.02) |
| | | KNN | 9.0<br>(8.0 , 11.5) | -2.14<br>(-2.26 , -2.03) | 0.87<br>(0.79 , 0.96) |
| | | FOREST | 9.0<br>(6.0 , 12.0) | -2.16<br>(-2.28 , -2.08) | 0.89<br>(0.79 , 1.00) |
| | | MICEpmm | 10.0<br>(8.0 , 12.0) | -2.16<br>(-2.26 , -2.08) | 0.88<br>(0.79 , 0.96) |
| | | MICEfcs | 10.0<br>(8.0 , 13.0) | -2.16<br>(-2.27 , -2.05) | 0.89<br>(0.78 , 0.99) |

**Supplementary Table S5.** Number of retained predictors in model and distribution of Linear predictor (LP)

| Missing | Scenario | Method | Retained Variables<br>Median (IQR) | Mean LP<br>Median (IQR) | SD LP<br>Median (IQR) |
|---|---|---|---|---|---|
| | LinMix | CCA | 7.0<br>(5.0 , 10.0) | -2.65<br>(-2.83 , -2.47) | 0.94<br>(0.77 , 1.08) |
| | | KNN | 9.0<br>(7.0 , 11.0) | -2.19<br>(-2.31 , -2.05) | 0.94<br>(0.85 , 1.06) |
| | | FOREST | 8.0<br>(7.0 , 12.0) | -2.19<br>(-2.38 , -2.09) | 0.96<br>(0.84 , 1.07) |
| | | MICEpmm | 10.0<br>(7.5 , 13.0) | -2.22<br>(-2.32 , -2.07) | 0.97<br>(0.86 , 1.06) |
| | | MICEfcs | 10.0<br>(8.0 , 13.5) | -2.22<br>(-2.33 , -2.12) | 0.97<br>(0.88 , 1.09) |
| | NLinMix | CCA | 8.0<br>(5.0 , 11.0) | -2.66<br>(-2.81 , -2.48) | 0.89<br>(0.76 , 1.08) |
| | | KNN | 9.0<br>(7.0 , 11.5) | -2.12<br>(-2.23 , -2.01) | 0.82<br>(0.72 , 0.93) |
| | | FOREST | 8.0<br>(7.0 , 11.0) | -2.13<br>(-2.25 , -2.03) | 0.87<br>(0.75 , 0.96) |
| | | MICEpmm | 10.0<br>(8.0 , 12.0) | -2.13<br>(-2.28 , -2.03) | 0.84<br>(0.74 , 0.96) |
| | | MICEfcs | 10.0<br>(8.0 , 12.5) | -2.14<br>(-2.25 , -2.03) | 0.84<br>(0.73 , 0.93) |
| | All Conditions | CCA | 9.0<br>(6.0 , 12.0) | -2.75<br>(-2.91 , -2.59) | 1.03<br>(0.86 , 1.18) |
| | | KNN | 9.0<br>(7.0 , 11.0) | -2.11<br>(-2.21 , -2.03) | 0.82<br>(0.74 , 0.91) |
| | | FOREST | 8.0<br>(6.0 , 12.0) | -2.17<br>(-2.27 , -2.07) | 0.91<br>(0.82 , 1.02) |
| | | MICEpmm | 10.0<br>(8.0 , 13.0) | -2.16<br>(-2.27 , -2.07) | 0.86<br>(0.79 , 0.95) |
| | | MICEfcs | 10.0<br>(8.0 , 12.0) | -2.16<br>(-2.28 , -2.06) | 0.88<br>(0.78 , 0.97) |

**Supplementary Table S5.** Number of retained predictors in model and distribution of Linear predictor (LP)

| Missing | Scenario | Method | Retained Variables Median (IQR) | Mean LP Median (IQR) | SD LP Median (IQR) |
|---|---|---|---|---|---|
| 60% Missing | Categorical | CCA | 6.0 (3.0 , 8.0) | -3.52 (-3.84 , -3.21) | 1.33 (1.01 , 1.68) |
| | | KNN | 8.0 (6.0 , 11.0) | -2.17 (-2.30 , -2.06) | 0.94 (0.85 , 1.06) |
| | | FOREST | 8.0 (6.0 , 11.0) | -2.20 (-2.34 , -2.10) | 0.97 (0.86 , 1.08) |
| | | MICEpmm | 9.0 (7.5 , 12.5) | -2.24 (-2.36 , -2.12) | 0.97 (0.86 , 1.09) |
| | | MICEfcs | 9.0 (7.0 , 12.5) | -2.24 (-2.37 , -2.10) | 1.00 (0.85 , 1.12) |
| | LinCon | CCA | 4.0 (1.5 , 7.0) | -3.37 (-3.63 , -3.08) | 0.90 (0.58 , 1.29) |
| | | KNN | 8.0 (7.0 , 11.0) | -2.20 (-2.31 , -2.12) | 0.97 (0.87 , 1.07) |
| | | FOREST | 8.0 (6.0 , 10.0) | -2.20 (-2.33 , -2.11) | 0.96 (0.89 , 1.06) |
| | | MICEpmm | 9.0 (8.0 , 12.0) | -2.21 (-2.33 , -2.12) | 0.97 (0.86 , 1.06) |
| | | MICEfcs | 10.0 (7.0 , 12.0) | -2.20 (-2.30 , -2.09) | 0.96 (0.87 , 1.08) |
| | NLinCon | CCA | 6.0 (4.0 , 10.0) | -3.50 (-3.87 , -3.18) | 1.26 (0.87 , 1.68) |
| | | KNN | 9.0 (7.5 , 12.0) | -2.16 (-2.27 , -2.03) | 0.90 (0.80 , 0.99) |
| | | FOREST | 10.0 (7.0 , 12.5) | -2.15 (-2.36 , -2.03) | 0.89 (0.79 , 1.01) |
| | | MICEpmm | 10.0 (8.0 , 12.5) | -2.18 (-2.31 , -2.07) | 0.89 (0.79 , 0.99) |
| | | MICEfcs | 10.0 (7.5 , 13.5) | -2.20 (-2.32 , -2.05) | 0.90 (0.79 , 1.01) |

**Supplementary Table S5.** Number of retained predictors in model and distribution of Linear predictor (LP)

| Missing | Scenario | Method | Retained Variables Median (IQR) | Mean LP Median (IQR) | SD LP Median (IQR) |
|---|---|---|---|---|---|
| | LinMix | CCA | 5.0 (2.0 , 9.0) | -2.95 (-3.28 , -2.68) | 0.95 (0.72 , 1.19) |
| | | KNN | 8.0 (5.0 , 11.0) | -2.17 (-2.33 , -2.08) | 0.91 (0.82 , 1.02) |
| | | FOREST | 8.0 (6.0 , 11.0) | -2.19 (-2.28 , -2.08) | 0.94 (0.83 , 1.05) |
| | | MICEpmm | 9.0 (7.0 , 12.0) | -2.20 (-2.32 , -2.09) | 0.93 (0.83 , 1.06) |
| | | MICEfcs | 9.0 (7.0 , 12.0) | -2.21 (-2.35 , -2.10) | 0.94 (0.84 , 1.04) |
| | NLinMix | CCA | 4.0 (2.5 , 8.0) | -2.99 (-3.30 , -2.69) | 0.92 (0.65 , 1.22) |
| | | KNN | 8.0 (6.0 , 10.0) | -2.17 (-2.30 , -2.03) | 0.91 (0.78 , 1.02) |
| | | FOREST | 8.0 (5.5 , 10.5) | -2.18 (-2.31 , -2.06) | 0.94 (0.84 , 1.04) |
| | | MICEpmm | 8.0 (7.0 , 11.0) | -2.19 (-2.32 , -2.08) | 0.92 (0.82 , 1.03) |
| | | MICEfcs | 9.0 (6.0 , 11.0) | -2.20 (-2.36 , -2.09) | 0.94 (0.83 , 1.05) |
| | All Conditions | CCA | 5.0 (2.5 , 8.5) | -3.01 (-3.32 , -2.77) | 1.01 (0.81 , 1.24) |
| | | KNN | 8.0 (6.0 , 10.5) | -2.10 (-2.22 , -2.00) | 0.80 (0.70 , 0.90) |
| | | FOREST | 8.0 (5.5 , 11.5) | -2.14 (-2.26 , -2.02) | 0.83 (0.74 , 0.97) |
| | | MICEpmm | 9.0 (6.0 , 12.0) | -2.12 (-2.24 , -2.03) | 0.80 (0.70 , 0.92) |
| | | MICEfcs | 9.0 (6.0 , 11.0) | -2.14 (-2.27 , -2.03) | 0.82 (0.72 , 0.92) |

**Supplementary Table S5.** Number of retained predictors in model and distribution of Linear predictor (LP)

| Missing | Scenario | Method | Retained Variables Median (IQR) | Mean LP Median (IQR) | SD LP Median (IQR) |
|---|---|---|---|---|---|
| 90% Missing | Categorical | CCA | 0.0 (0.0 , 2.0) | -2.94 (-3.59 , -2.60) | 0.00 (0.00 , 1.32) |
| | | KNN | 8.0 (5.5 , 11.0) | -2.17 (-2.26 , -2.07) | 0.91 (0.81 , 1.01) |
| | | FOREST | 8.0 (6.0 , 10.5) | -2.17 (-2.29 , -2.06) | 0.91 (0.80 , 0.99) |
| | | MICEpmm | 9.0 (6.5 , 11.0) | -2.18 (-2.30 , -2.06) | 0.91 (0.84 , 1.06) |
| | | MICEfcs | 9.0 (5.5 , 12.0) | -2.19 (-2.39 , -2.06) | 0.93 (0.82 , 1.06) |
| | LinCon | CCA | 0.0 (0.0 , 1.0) | -3.50 (-4.11 , -3.11) | 0.00 (0.00 , 1.25) |
| | | KNN | 8.0 (6.5 , 11.0) | -2.19 (-2.29 , -2.07) | 0.96 (0.87 , 1.04) |
| | | FOREST | 8.0 (7.0 , 11.0) | -2.21 (-2.31 , -2.10) | 0.96 (0.87 , 1.05) |
| | | MICEpmm | 9.0 (7.0 , 11.5) | -2.20 (-2.37 , -2.08) | 0.97 (0.85 , 1.11) |
| | | MICEfcs | 9.0 (6.0 , 11.0) | -2.22 (-2.40 , -2.13) | 0.97 (0.87 , 1.12) |
| | NLinCon | CCA | 0.0 (0.0 , 4.0) | -3.33 (-4.14 , -2.88) | 0.00 (0.00 , 1.74) |
| | | KNN | 9.0 (6.0 , 12.0) | -2.14 (-2.25 , -2.02) | 0.86 (0.73 , 0.98) |
| | | FOREST | 9.0 (6.0 , 12.0) | -2.13 (-2.30 , -2.01) | 0.87 (0.76 , 0.98) |
| | | MICEpmm | 9.0 (6.0 , 11.5) | -2.19 (-2.45 , -2.02) | 0.89 (0.77 , 1.08) |
| | | MICEfcs | 8.0 (6.0 , 11.0) | -2.19 (-2.70 , -2.05) | 0.87 (0.77 , 1.04) |

**Supplementary Table S5.** Number of retained predictors in model and distribution of Linear predictor (LP)

| Missing | Scenario | Method | Retained Variables Median (IQR) | Mean LP Median (IQR) | SD LP Median (IQR) |
|---|---|---|---|---|---|
| | LinMix | CCA | 0.0 (0.0 , 2.5) | -3.20 (-3.84 , -2.80) | 0.00 (0.00 , 1.38) |
| | | KNN | 7.5 (5.0 , 10.5) | -2.17 (-2.28 , -2.06) | 0.91 (0.80 , 1.01) |
| | | FOREST | 7.0 (5.5 , 9.5) | -2.18 (-2.28 , -2.05) | 0.91 (0.81 , 1.00) |
| | | MICEpmm | 8.0 (5.5 , 10.5) | -2.18 (-2.38 , -2.07) | 0.90 (0.82 , 1.05) |
| | | MICEfcs | 8.0 (6.0 , 10.5) | -2.16 (-2.59 , -2.07) | 0.92 (0.81 , 1.13) |
| | NLinMix | CCA | 1.0 (0.0 , 4.0) | -3.08 (-3.62 , -2.67) | 0.37 (0.00 , 2.04) |
| | | KNN | 8.0 (6.0 , 11.0) | -2.11 (-2.23 , -2.01) | 0.81 (0.71 , 0.87) |
| | | FOREST | 8.0 (5.5 , 10.5) | -2.11 (-2.27 , -1.97) | 0.81 (0.70 , 0.94) |
| | | MICEpmm | 8.0 (5.5 , 11.0) | -2.13 (-2.35 , -2.00) | 0.81 (0.72 , 0.97) |
| | | MICEfcs | 8.0 (5.0 , 11.0) | -2.13 (-2.47 , -2.01) | 0.81 (0.70 , 0.95) |
| | All Conditions | CCA | 0.0 (0.0 , 4.0) | -3.27 (-4.00 , -2.91) | 0.00 (0.00 , 1.90) |
| | | KNN | 8.0 (4.5 , 12.0) | -2.10 (-2.23 , -1.98) | 0.78 (0.67 , 0.90) |
| | | FOREST | 7.0 (5.0 , 10.0) | -2.11 (-2.25 , -2.02) | 0.79 (0.69 , 0.89) |
| | | MICEpmm | 7.5 (5.0 , 10.0) | -2.21 (-2.91 , -2.02) | 0.86 (0.70 , 1.12) |
| | | MICEfcs | 7.0 (4.0 , 9.0) | -2.25 (-3.09 , -2.01) | 0.83 (0.68 , 1.16) |

**Supplementary Table S6.** Optimism-corrected performance, prediction stability, and computational time.

| Missing | Scenario | Method | Corrected AUC | Calibration Slope | Proportion MAPE ≤5% | Time (Sec) |
|---|---|---|---|---|---|---|
| 30% Missing | Categorical | CCA | 0.75 (0.73 , 0.76) | 0.93 (0.82 , 1.06) | 0.64 (0.57 , 0.69) | 0.6 (0.6 , 0.8) |
| | | KNN | 0.75 (0.73 , 0.75) | 0.96 (0.89 , 1.08) | 0.90 (0.82 , 0.97) | 8.3 (8.0 , 8.8) |
| | | FOREST | 0.75 (0.74 , 0.75) | 0.96 (0.87 , 1.09) | 0.91 (0.83 , 0.98) | 7.2 (6.5 , 17.2) |
| | | MICEpmm | 0.75 (0.74 , 0.75) | 0.96 (0.89 , 1.05) | 0.93 (0.86 , 0.98) | 61.9 (58.0 , 67.4) |
| | | MICEfcs | 0.75 (0.74 , 0.75) | 0.96 (0.89 , 1.06) | 0.92 (0.87 , 0.98) | 66.1 (61.8 , 71.1) |
| | LinCon | CCA | 0.75 (0.74 , 0.76) | 0.93 (0.83 , 1.07) | 0.63 (0.59 , 0.69) | 0.6 (0.6 , 0.9) |
| | | KNN | 0.75 (0.74 , 0.75) | 0.96 (0.89 , 1.05) | 0.91 (0.83 , 0.98) | 8.2 (7.8 , 8.4) |
| | | FOREST | 0.75 (0.74 , 0.75) | 0.97 (0.89 , 1.07) | 0.92 (0.85 , 0.99) | 34.5 (16.9 , 91.6) |
| | | MICEpmm | 0.75 (0.75 , 0.75) | 0.95 (0.88 , 1.06) | 0.93 (0.86 , 0.99) | 60.5 (56.0 , 66.4) |
| | | MICEfcs | 0.75 (0.74 , 0.76) | 0.96 (0.86 , 1.05) | 0.92 (0.86 , 0.97) | 64.8 (56.7 , 72.1) |
| | NLinCon | CCA | 0.72 (0.70 , 0.73) | 0.91 (0.79 , 1.04) | 0.64 (0.59 , 0.69) | 0.6 (0.6 , 1.0) |
| | | KNN | 0.74 (0.73 , 0.74) | 0.98 (0.88 , 1.06) | 0.90 (0.83 , 0.99) | 23.1 (22.4 , 23.7) |
| | | FOREST | 0.73 (0.72 , 0.74) | 0.95 (0.85 , 1.05) | 0.90 (0.83 , 0.96) | 56.9 (34.0 , 125.8) |
| | | MICEpmm | 0.74 (0.73 , 0.75) | 0.95 (0.87 , 1.05) | 0.92 (0.86 , 0.97) | 61.1 (57.4 , 66.0) |
| | | MICEfcs | 0.74 (0.73 , 0.74) | 0.95 (0.85 , 1.06) | 0.92 (0.86 , 0.97) | 73.4 (59.7 , 81.2) |

**Supplementary Table S6.** Optimism-corrected performance, prediction stability, and computational time.

| Missing | Scenario | Method | Corrected AUC | Calibration Slope | Proportion MAPE ≤5% | Time (Sec) |
|---|---|---|---|---|---|---|
| | LinMix | CCA | 0.73 (0.72 , 0.74) | 0.92 (0.82 , 1.10) | 0.61 (0.57 , 0.66) | 0.6 (0.6 , 0.9) |
| | | KNN | 0.75 (0.74 , 0.75) | 0.97 (0.88 , 1.07) | 0.89 (0.82 , 0.97) | 16.6 (16.1 , 17.0) |
| | | FOREST | 0.75 (0.74 , 0.75) | 0.97 (0.88 , 1.05) | 0.90 (0.82 , 0.97) | 43.4 (20.2 , 117.0) |
| | | MICEpmm | 0.75 (0.74 , 0.76) | 0.95 (0.85 , 1.05) | 0.91 (0.85 , 0.97) | 60.1 (54.7 , 64.8) |
| | | MICEfcs | 0.75 (0.74 , 0.76) | 0.95 (0.87 , 1.03) | 0.91 (0.84 , 0.97) | 67.6 (59.5 , 73.6) |
| | NLinMix | CCA | 0.72 (0.71 , 0.74) | 0.90 (0.79 , 1.04) | 0.61 (0.57 , 0.67) | 0.6 (0.6 , 0.8) |
| | | KNN | 0.72 (0.71 , 0.73) | 0.95 (0.85 , 1.06) | 0.90 (0.81 , 0.98) | 31.7 (30.8 , 32.5) |
| | | FOREST | 0.73 (0.72 , 0.74) | 0.95 (0.87 , 1.09) | 0.91 (0.83 , 0.98) | 61.9 (40.4 , 126.1) |
| | | MICEpmm | 0.73 (0.72 , 0.74) | 0.94 (0.84 , 1.04) | 0.91 (0.84 , 0.96) | 62.7 (58.2 , 66.3) |
| | | MICEfcs | 0.73 (0.72 , 0.74) | 0.93 (0.84 , 1.06) | 0.91 (0.86 , 0.96) | 69.8 (64.1 , 74.1) |
| | All Conditions | CCA | 0.76 (0.75 , 0.77) | 0.93 (0.84 , 1.05) | 0.63 (0.58 , 0.67) | 0.6 (0.5 , 0.8) |
| | | KNN | 0.73 (0.72 , 0.74) | 0.96 (0.88 , 1.07) | 0.90 (0.82 , 0.97) | 37.3 (36.4 , 39.4) |
| | | FOREST | 0.74 (0.73 , 0.74) | 0.95 (0.87 , 1.07) | 0.91 (0.83 , 0.97) | 78.4 (49.3 , 93.6) |
| | | MICEpmm | 0.74 (0.73 , 0.75) | 0.95 (0.85 , 1.05) | 0.89 (0.81 , 0.96) | 63.8 (58.7 , 67.9) |
| | | MICEfcs | 0.74 (0.73 , 0.75) | 0.94 (0.85 , 1.07) | 0.91 (0.86 , 0.95) | 67.0 (63.1 , 71.2) |

**Supplementary Table S6.** Optimism-corrected performance, prediction stability, and computational time.

| Missing | Scenario | Method | Corrected AUC | Calibration Slope | Proportion MAPE ≤5% | Time (Sec) |
|---|---|---|---|---|---|---|
| 60% Missing | Categorical | CCA | 0.80 (0.77 , 0.82) | 0.90 (0.77 , 1.06) | 0.36 (0.34 , 0.39) | 0.4 (0.4 , 0.7) |
| | | KNN | 0.75 (0.74 , 0.75) | 0.96 (0.87 , 1.06) | 0.89 (0.81 , 0.97) | 10.2 (9.8 , 10.5) |
| | | FOREST | 0.75 (0.74 , 0.75) | 0.96 (0.86 , 1.08) | 0.90 (0.84 , 0.97) | 29.0 (6.3 , 82.3) |
| | | MICEpmm | 0.75 (0.74 , 0.76) | 0.95 (0.85 , 1.06) | 0.90 (0.85 , 0.95) | 62.7 (57.0 , 66.1) |
| | | MICEfcs | 0.76 (0.74 , 0.76) | 0.94 (0.85 , 1.04) | 0.91 (0.84 , 0.96) | 70.5 (65.0 , 75.8) |
| | LinCon | CCA | 0.69 (0.64 , 0.72) | 0.82 (0.66 , 1.08) | 0.38 (0.35 , 0.42) | 0.4 (0.4 , 0.7) |
| | | KNN | 0.75 (0.75 , 0.75) | 0.96 (0.87 , 1.05) | 0.92 (0.83 , 0.99) | 10.2 (9.9 , 10.5) |
| | | FOREST | 0.75 (0.74 , 0.75) | 0.97 (0.89 , 1.06) | 0.93 (0.85 , 0.99) | 26.3 (10.1 , 85.1) |
| | | MICEpmm | 0.75 (0.74 , 0.75) | 0.96 (0.86 , 1.08) | 0.93 (0.83 , 0.98) | 61.2 (57.1 , 66.0) |
| | | MICEfcs | 0.75 (0.74 , 0.75) | 0.97 (0.87 , 1.08) | 0.94 (0.85 , 0.99) | 70.9 (63.7 , 76.3) |
| | NLinCon | CCA | 0.76 (0.74 , 0.79) | 0.85 (0.72 , 1.05) | 0.36 (0.33 , 0.39) | 0.4 (0.4 , 0.7) |
| | | KNN | 0.74 (0.73 , 0.75) | 0.95 (0.86 , 1.09) | 0.91 (0.82 , 0.99) | 27.7 (27.3 , 28.6) |
| | | FOREST | 0.73 (0.72 , 0.74) | 0.96 (0.83 , 1.06) | 0.89 (0.83 , 0.95) | 36.3 (18.4 , 111.5) |
| | | MICEpmm | 0.75 (0.72 , 0.76) | 0.93 (0.81 , 1.03) | 0.89 (0.82 , 0.96) | 61.9 (56.8 , 66.1) |
| | | MICEfcs | 0.75 (0.73 , 0.75) | 0.93 (0.80 , 1.07) | 0.88 (0.78 , 0.93) | 69.9 (60.1 , 77.7) |

**Supplementary Table S6.** Optimism-corrected performance, prediction stability, and computational time.

| Missing | Scenario | Method | Corrected AUC | Calibration Slope | Proportion MAPE ≤5% | Time (Sec) |
|---|---|---|---|---|---|---|
| | LinMix | CCA | 0.73 (0.71 , 0.74) | 0.87 (0.73 , 1.05) | 0.35 (0.32 , 0.39) | 0.4 (0.4 , 0.7) |
| | | KNN | 0.74 (0.73 , 0.75) | 0.96 (0.88 , 1.06) | 0.90 (0.81 , 0.97) | 19.3 (18.7 , 19.9) |
| | | FOREST | 0.74 (0.74 , 0.75) | 0.95 (0.88 , 1.04) | 0.91 (0.83 , 0.97) | 32.0 (12.7 , 99.1) |
| | | MICEpmm | 0.74 (0.74 , 0.75) | 0.94 (0.84 , 1.06) | 0.91 (0.85 , 0.97) | 63.1 (57.9 , 67.9) |
| | | MICEfcs | 0.74 (0.73 , 0.75) | 0.94 (0.84 , 1.05) | 0.91 (0.84 , 0.97) | 70.5 (66.2 , 76.4) |
| | NLinMix | CCA | 0.71 (0.68 , 0.72) | 0.86 (0.67 , 1.11) | 0.35 (0.30 , 0.39) | 0.4 (0.3 , 0.6) |
| | | KNN | 0.74 (0.73 , 0.74) | 0.96 (0.85 , 1.08) | 0.92 (0.85 , 0.98) | 19.0 (18.6 , 19.4) |
| | | FOREST | 0.74 (0.73 , 0.75) | 0.96 (0.86 , 1.08) | 0.87 (0.81 , 0.94) | 28.8 (11.3 , 89.4) |
| | | MICEpmm | 0.74 (0.73 , 0.75) | 0.95 (0.83 , 1.06) | 0.91 (0.85 , 0.98) | 64.2 (60.7 , 69.7) |
| | | MICEfcs | 0.74 (0.73 , 0.75) | 0.93 (0.83 , 1.03) | 0.89 (0.79 , 0.98) | 81.9 (64.9 , 89.8) |
| | All Conditions | CCA | 0.73 (0.71 , 0.74) | 0.88 (0.75 , 1.05) | 0.34 (0.31 , 0.38) | 0.4 (0.4 , 0.6) |
| | | KNN | 0.71 (0.71 , 0.72) | 0.92 (0.83 , 1.05) | 0.88 (0.81 , 0.95) | 44.4 (43.9 , 46.2) |
| | | FOREST | 0.72 (0.71 , 0.73) | 0.94 (0.83 , 1.05) | 0.89 (0.80 , 0.97) | 41.0 (26.0 , 122.3) |
| | | MICEpmm | 0.72 (0.71 , 0.73) | 0.92 (0.81 , 1.06) | 0.89 (0.80 , 0.94) | 66.4 (62.3 , 70.0) |
| | | MICEfcs | 0.72 (0.70 , 0.73) | 0.91 (0.81 , 1.03) | 0.88 (0.82 , 0.95) | 75.2 (68.0 , 80.3) |

**Supplementary Table S6.** Optimism-corrected performance, prediction stability, and computational time.

| Missing | Scenario | Method | Corrected AUC | Calibration Slope | Proportion MAPE ≤5% | Time (Sec) |
|---|---|---|---|---|---|---|
| 90% Missing | Categorical | CCA | 0.50 (0.50 , 0.67) | 0.69 (0.27 , 1.12) | 0.10 (0.06 , 0.10) | 0.2 (0.2 , 0.2) |
| | | KNN | 0.74 (0.73 , 0.75) | 0.96 (0.89 , 1.09) | 0.90 (0.84 , 0.98) | 5.5 (5.2 , 6.0) |
| | | FOREST | 0.74 (0.73 , 0.74) | 0.96 (0.89 , 1.10) | 0.91 (0.79 , 0.99) | 24.8 (5.8 , 87.9) |
| | | MICEpmm | 0.74 (0.73 , 0.75) | 0.96 (0.80 , 1.05) | 0.92 (0.83 , 0.98) | 60.9 (54.4 , 66.2) |
| | | MICEfcs | 0.74 (0.73 , 0.75) | 0.95 (0.78 , 1.05) | 0.91 (0.82 , 0.97) | 70.5 (63.9 , 78.9) |
| | LinCon | CCA | 0.50 (0.50 , 0.73) | 0.79 (0.42 , 0.95) | 0.11 (0.09 , 0.11) | 0.2 (0.2 , 0.4) |
| | | KNN | 0.75 (0.75 , 0.75) | 0.97 (0.90 , 1.08) | 0.92 (0.82 , 1.00) | 5.2 (4.9 , 5.6) |
| | | FOREST | 0.75 (0.74 , 0.75) | 0.96 (0.89 , 1.06) | 0.93 (0.84 , 0.99) | 32.6 (6.0 , 93.8) |
| | | MICEpmm | 0.75 (0.74 , 0.75) | 0.97 (0.80 , 1.05) | 0.93 (0.82 , 0.99) | 61.7 (57.5 , 66.3) |
| | | MICEfcs | 0.75 (0.74 , 0.75) | 0.96 (0.77 , 1.06) | 0.92 (0.82 , 0.99) | 67.2 (63.1 , 73.4) |
| | NLinCon | CCA | 0.50 (0.50 , 0.77) | 0.73 (0.25 , 1.11) | 0.10 (0.08 , 0.10) | 0.2 (0.2 , 0.4) |
| | | KNN | 0.73 (0.72 , 0.74) | 0.95 (0.84 , 1.08) | 0.92 (0.83 , 0.97) | 13.9 (13.2 , 14.5) |
| | | FOREST | 0.73 (0.72 , 0.74) | 0.94 (0.85 , 1.06) | 0.92 (0.82 , 0.99) | 30.4 (7.1 , 97.6) |
| | | MICEpmm | 0.73 (0.68 , 0.76) | 0.91 (0.49 , 1.04) | 0.88 (0.67 , 0.98) | 64.3 (60.1 , 68.6) |
| | | MICEfcs | 0.73 (0.65 , 0.75) | 0.90 (0.39 , 1.05) | 0.86 (0.68 , 0.98) | 69.9 (64.2 , 74.7) |

**Supplementary Table S6.** Optimism-corrected performance, prediction stability, and computational time.

| Missing | Scenario | Method | Corrected AUC | Calibration Slope | Proportion MAPE ≤5% | Time (Sec) |
|---|---|---|---|---|---|---|
| | LinMix | CCA | 0.50 (0.50 , 0.64) | 0.56 (0.13 , 0.88) | 0.09 (0.06 , 0.09) | 0.3 (0.3 , 0.4) |
| | | KNN | 0.74 (0.73 , 0.74) | 0.96 (0.87 , 1.07) | 0.92 (0.83 , 0.98) | 9.5 (9.1 , 10.1) |
| | | FOREST | 0.74 (0.73 , 0.74) | 0.96 (0.88 , 1.06) | 0.93 (0.84 , 0.99) | 27.6 (6.5 , 96.7) |
| | | MICEpmm | 0.74 (0.72 , 0.75) | 0.95 (0.71 , 1.06) | 0.93 (0.83 , 0.99) | 64.6 (61.7 , 68.6) |
| | | MICEfcs | 0.74 (0.70 , 0.74) | 0.95 (0.56 , 1.09) | 0.93 (0.72 , 0.99) | 75.6 (66.4 , 81.6) |
| | NLinMix | CCA | 0.54 (0.50 , 0.67) | 0.58 (0.27 , 0.87) | 0.10 (0.07 , 0.10) | 0.2 (0.2 , 0.3) |
| | | KNN | 0.72 (0.71 , 0.72) | 0.96 (0.88 , 1.09) | 0.91 (0.83 , 0.97) | 18.6 (17.8 , 19.2) |
| | | FOREST | 0.72 (0.71 , 0.72) | 0.96 (0.83 , 1.09) | 0.90 (0.80 , 0.97) | 24.4 (7.5 , 90.6) |
| | | MICEpmm | 0.72 (0.68 , 0.72) | 0.94 (0.58 , 1.07) | 0.93 (0.69 , 0.99) | 67.7 (63.5 , 72.6) |
| | | MICEfcs | 0.72 (0.66 , 0.72) | 0.94 (0.55 , 1.08) | 0.92 (0.69 , 0.98) | 75.1 (68.9 , 80.8) |
| | All Conditions | CCA | 0.50 (0.50 , 0.81) | 0.77 (0.40 , 1.07) | 0.08 (0.06 , 0.09) | 0.2 (0.2 , 0.4) |
| | | KNN | 0.71 (0.70 , 0.72) | 0.95 (0.84 , 1.09) | 0.90 (0.82 , 0.95) | 21.9 (21.2 , 22.8) |
| | | FOREST | 0.71 (0.70 , 0.72) | 0.94 (0.82 , 1.05) | 0.91 (0.82 , 0.99) | 37.1 (7.6 , 85.2) |
| | | MICEpmm | 0.71 (0.63 , 0.74) | 0.83 (0.32 , 1.04) | 0.85 (0.61 , 0.98) | 67.4 (61.7 , 73.0) |
| | | MICEfcs | 0.71 (0.62 , 0.74) | 0.83 (0.24 , 1.08) | 0.84 (0.62 , 1.00) | 77.4 (72.0 , 81.9) |

**Supplementary Table S7.** External validation of predictive performance.

| Missing | Scenario | Method | External AUC | External Cal Slope |
|---|---|---|---|---|
| 30% Missing | Categorical | CCA | 0.71 (0.70 , 0.73) | 0.77 (0.66 , 0.92) |
| | | KNN | 0.72 (0.71 , 0.73) | 0.84 (0.77 , 0.94) |
| | | FOREST | 0.72 (0.70 , 0.73) | 0.89 (0.77 , 1.07) |
| | | MICEpmm | 0.72 (0.70 , 0.73) | 0.92 (0.80 , 1.07) |
| | | MICEfcs | 0.72 (0.70 , 0.73) | 0.93 (0.80 , 1.05) |
| | LinCon | CCA | 0.71 (0.69 , 0.72) | 0.75 (0.61 , 0.88) |
| | | KNN | 0.72 (0.72 , 0.73) | 0.83 (0.76 , 0.90) |
| | | FOREST | 0.72 (0.71 , 0.73) | 0.87 (0.79 , 1.02) |
| | | MICEpmm | 0.72 (0.70 , 0.73) | 0.89 (0.78 , 0.99) |
| | | MICEfcs | 0.72 (0.70 , 0.73) | 0.89 (0.74 , 1.04) |
| | NLinCon | CCA | 0.71 (0.68 , 0.72) | 0.82 (0.64 , 0.97) |
| | | KNN | 0.70 (0.68 , 0.71) | 0.86 (0.75 , 0.96) |
| | | FOREST | 0.59 (0.54 , 0.65) | 0.41 (0.17 , 0.71) |
| | | MICEpmm | 0.66 (0.60 , 0.69) | 0.76 (0.50 , 0.96) |
| | | MICEfcs | 0.65 (0.60 , 0.69) | 0.72 (0.47 , 0.90) |

**Supplementary Table S7.** External validation of predictive performance.

| Missing | Scenario | Method | External AUC | External Cal Slope |
|---|---|---|---|---|
| | LinMix | CCA | 0.72 (0.70 , 0.73) | 0.79 (0.66 , 0.97) |
| | | KNN | 0.72 (0.72 , 0.73) | 0.84 (0.77 , 0.93) |
| | | FOREST | 0.72 (0.70 , 0.73) | 0.88 (0.77 , 1.00) |
| | | MICEpmm | 0.72 (0.70 , 0.73) | 0.87 (0.76 , 0.99) |
| | | MICEfcs | 0.72 (0.69 , 0.73) | 0.86 (0.75 , 0.99) |
| | NLinMix | CCA | 0.71 (0.70 , 0.73) | 0.82 (0.69 , 0.97) |
| | | KNN | 0.70 (0.68 , 0.72) | 0.88 (0.77 , 1.00) |
| | | FOREST | 0.63 (0.55 , 0.67) | 0.66 (0.27 , 0.85) |
| | | MICEpmm | 0.66 (0.61 , 0.69) | 0.81 (0.56 , 1.03) |
| | | MICEfcs | 0.66 (0.57 , 0.70) | 0.81 (0.39 , 1.05) |
| | All Conditions | CCA | 0.71 (0.69 , 0.72) | 0.70 (0.57 , 0.82) |
| | | KNN | 0.70 (0.68 , 0.71) | 0.88 (0.79 , 0.98) |
| | | FOREST | 0.65 (0.59 , 0.69) | 0.70 (0.42 , 0.87) |
| | | MICEpmm | 0.66 (0.63 , 0.69) | 0.79 (0.59 , 0.97) |
| | | MICEfcs | 0.66 (0.62 , 0.69) | 0.76 (0.54 , 0.96) |

**Supplementary Table S7.** External validation of predictive performance.

| Missing | Scenario | Method | External AUC | External Cal Slope |
|---|---|---|---|---|
| 60% Missing | Categorical | CCA | 0.70 (0.68 , 0.72) | 0.55 (0.44 , 0.65) |
| | | KNN | 0.72 (0.71 , 0.73) | 0.84 (0.75 , 0.92) |
| | | FOREST | 0.72 (0.70 , 0.73) | 0.85 (0.74 , 1.02) |
| | | MICEpmm | 0.72 (0.70 , 0.73) | 0.84 (0.72 , 0.98) |
| | | MICEfcs | 0.71 (0.70 , 0.73) | 0.83 (0.68 , 1.02) |
| | LinCon | CCA | 0.68 (0.61 , 0.71) | 0.71 (0.39 , 1.10) |
| | | KNN | 0.72 (0.72 , 0.73) | 0.82 (0.73 , 0.90) |
| | | FOREST | 0.72 (0.70 , 0.73) | 0.88 (0.78 , 1.03) |
| | | MICEpmm | 0.72 (0.70 , 0.73) | 0.92 (0.76 , 1.02) |
| | | MICEfcs | 0.72 (0.70 , 0.73) | 0.91 (0.77 , 1.04) |
| | NLinCon | CCA | 0.68 (0.63 , 0.71) | 0.51 (0.26 , 0.72) |
| | | KNN | 0.68 (0.65 , 0.70) | 0.76 (0.62 , 0.91) |
| | | FOREST | 0.56 (0.52 , 0.60) | 0.27 (0.02 , 0.46) |
| | | MICEpmm | 0.66 (0.55 , 0.69) | 0.76 (0.28 , 0.95) |
| | | MICEfcs | 0.67 (0.55 , 0.70) | 0.73 (0.29 , 0.91) |

**Supplementary Table S7.** External validation of predictive performance.

| Missing | Scenario | Method | External AUC | External Cal Slope |
|---|---|---|---|---|
| | LinMix | CCA | 0.70 (0.68 , 0.72) | 0.71 (0.57 , 0.88) |
| | | KNN | 0.72 (0.71 , 0.72) | 0.85 (0.76 , 0.93) |
| | | FOREST | 0.72 (0.70 , 0.73) | 0.89 (0.77 , 1.05) |
| | | MICEpmm | 0.71 (0.69 , 0.73) | 0.88 (0.72 , 1.02) |
| | | MICEfcs | 0.71 (0.70 , 0.73) | 0.89 (0.74 , 1.02) |
| | NLinMix | CCA | 0.67 (0.62 , 0.70) | 0.64 (0.39 , 0.99) |
| | | KNN | 0.72 (0.71 , 0.72) | 0.86 (0.75 , 0.96) |
| | | FOREST | 0.71 (0.69 , 0.72) | 0.91 (0.78 , 1.04) |
| | | MICEpmm | 0.71 (0.68 , 0.72) | 0.89 (0.73 , 1.04) |
| | | MICEfcs | 0.71 (0.67 , 0.72) | 0.88 (0.66 , 1.02) |
| | All Conditions | CCA | 0.70 (0.67 , 0.71) | 0.68 (0.53 , 0.86) |
| | | KNN | 0.68 (0.65 , 0.70) | 0.83 (0.69 , 0.98) |
| | | FOREST | 0.61 (0.53 , 0.67) | 0.52 (0.17 , 0.85) |
| | | MICEpmm | 0.60 (0.52 , 0.67) | 0.48 (0.15 , 0.88) |
| | | MICEfcs | 0.58 (0.51 , 0.64) | 0.39 (0.11 , 0.75) |

**Supplementary Table S7.** External validation of predictive performance.

| Missing | Scenario | Method | External AUC | External Cal Slope |
|---|---|---|---|---|
| 90% Missing | Categorical | CCA | 0.50 (0.50 , 0.69) | 0.72 (0.02 , 1.08) |
| | | KNN | 0.72 (0.71 , 0.73) | 0.84 (0.77 , 0.97) |
| | | FOREST | 0.71 (0.69 , 0.73) | 0.87 (0.78 , 1.08) |
| | | MICEpmm | 0.71 (0.70 , 0.73) | 0.88 (0.68 , 1.04) |
| | | MICEfcs | 0.71 (0.69 , 0.72) | 0.87 (0.65 , 1.02) |
| | LinCon | CCA | 0.50 (0.50 , 0.69) | 0.52 (-0.01 , 0.76) |
| | | KNN | 0.72 (0.72 , 0.73) | 0.83 (0.76 , 0.92) |
| | | FOREST | 0.72 (0.70 , 0.73) | 0.90 (0.74 , 1.03) |
| | | MICEpmm | 0.72 (0.68 , 0.73) | 0.88 (0.59 , 1.02) |
| | | MICEfcs | 0.72 (0.66 , 0.73) | 0.89 (0.53 , 1.02) |
| | NLinCon | CCA | 0.50 (0.50 , 0.63) | 0.17 (0.04 , 0.49) |
| | | KNN | 0.68 (0.62 , 0.71) | 0.78 (0.53 , 0.94) |
| | | FOREST | 0.56 (0.48 , 0.60) | 0.28 (-0.12 , 0.48) |
| | | MICEpmm | 0.58 (0.51 , 0.70) | 0.36 (0.03 , 0.78) |
| | | MICEfcs | 0.58 (0.53 , 0.70) | 0.38 (0.11 , 0.83) |

**Supplementary Table S7.** External validation of predictive performance.

| Missing | Scenario | Method | External AUC | External Cal Slope |
|---|---|---|---|---|
| | LinMix | CCA | 0.50 (0.50 , 0.69) | 0.51 (-0.03 , 0.99) |
| | | KNN | 0.71 (0.70 , 0.72) | 0.83 (0.73 , 0.93) |
| | | FOREST | 0.71 (0.70 , 0.72) | 0.87 (0.78 , 1.04) |
| | | MICEpmm | 0.71 (0.68 , 0.72) | 0.90 (0.56 , 1.08) |
| | | MICEfcs | 0.71 (0.63 , 0.72) | 0.89 (0.40 , 1.11) |
| | NLinMix | CCA | 0.50 (0.49 , 0.69) | 0.48 (-0.10 , 0.81) |
| | | KNN | 0.68 (0.63 , 0.70) | 0.80 (0.60 , 0.97) |
| | | FOREST | 0.56 (0.52 , 0.68) | 0.32 (0.03 , 0.90) |
| | | MICEpmm | 0.55 (0.49 , 0.59) | 0.26 (0.00 , 0.43) |
| | | MICEfcs | 0.55 (0.51 , 0.60) | 0.24 (0.01 , 0.44) |
| | All Conditions | CCA | 0.50 (0.50 , 0.68) | 0.37 (0.10 , 0.69) |
| | | KNN | 0.68 (0.60 , 0.70) | 0.83 (0.45 , 0.95) |
| | | FOREST | 0.53 (0.43 , 0.60) | 0.18 (-0.27 , 0.46) |
| | | MICEpmm | 0.55 (0.46 , 0.63) | 0.21 (-0.14 , 0.49) |
| | | MICEfcs | 0.55 (0.49 , 0.67) | 0.23 (-0.03 , 0.70) |

**Supplementary Figure S1.** Instability plot and mean absolute predictor error (MAPE) plot across all missingness proportions in missing of single categorical scenarios

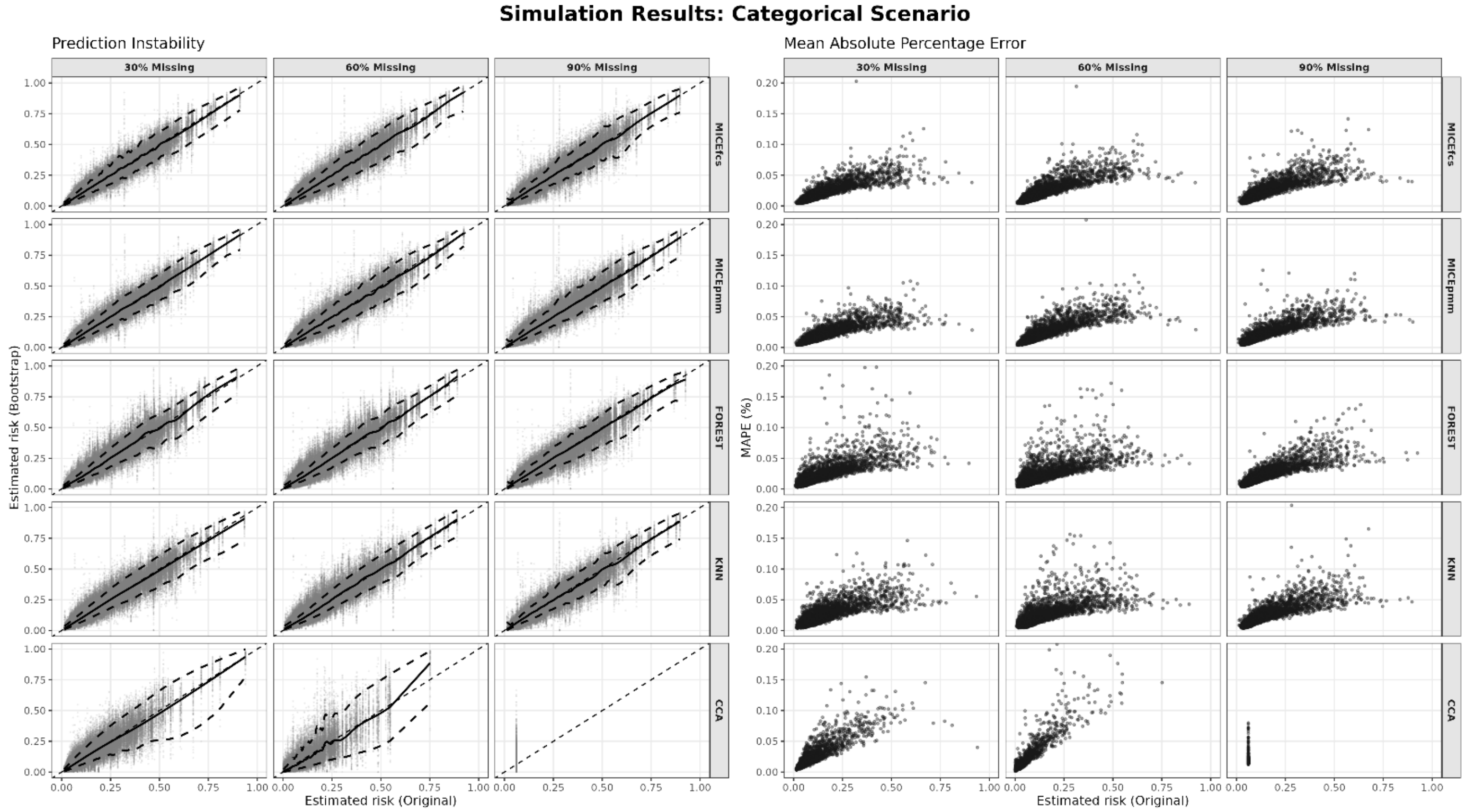

**Supplementary Figure S2.** Instability plot and mean absolute predictor error (MAPE) plot across all missingness proportions in missing of single linear continuous (LinCon) scenarios

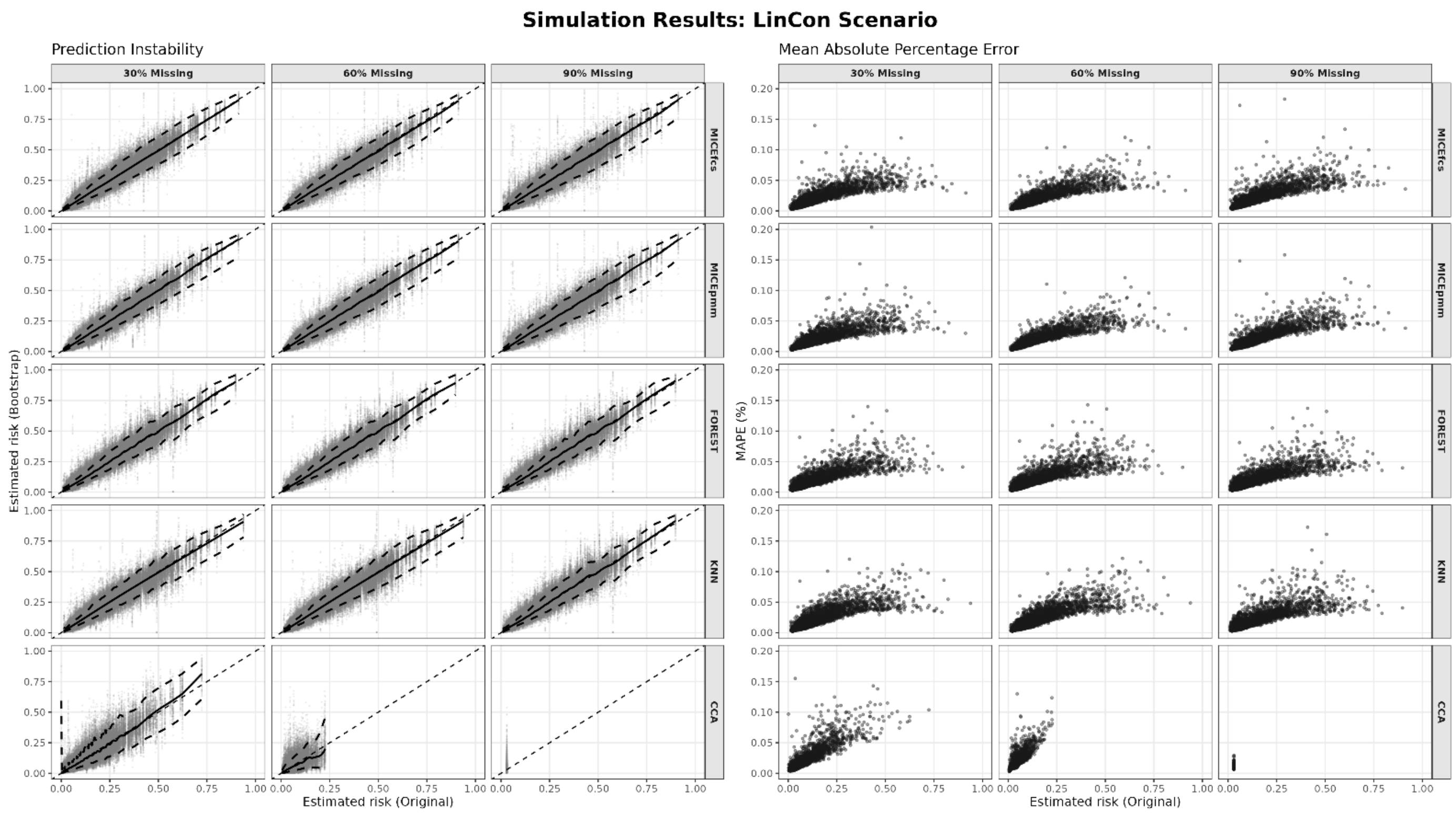

**Supplementary Figure S3.** Instability plot and mean absolute predictor error (MAPE) plot across all missingness proportions in missing of linear continuous and categorical (LinMix) scenarios

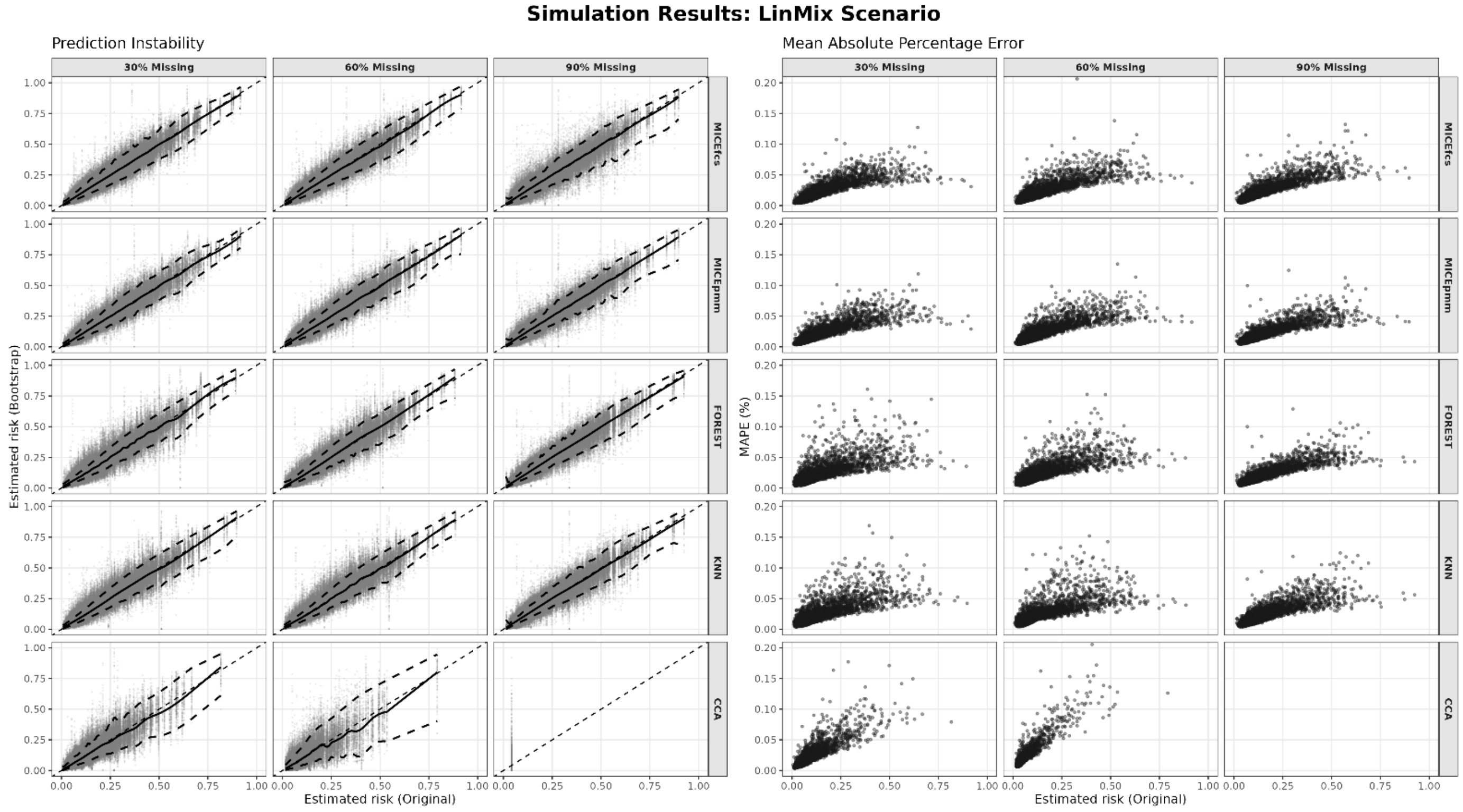

**Supplementary Figure S4.** Instability plot and mean absolute predictor error (MAPE) plot across all missingness proportions in missing of single non-linear continuous (NLinCon) scenarios

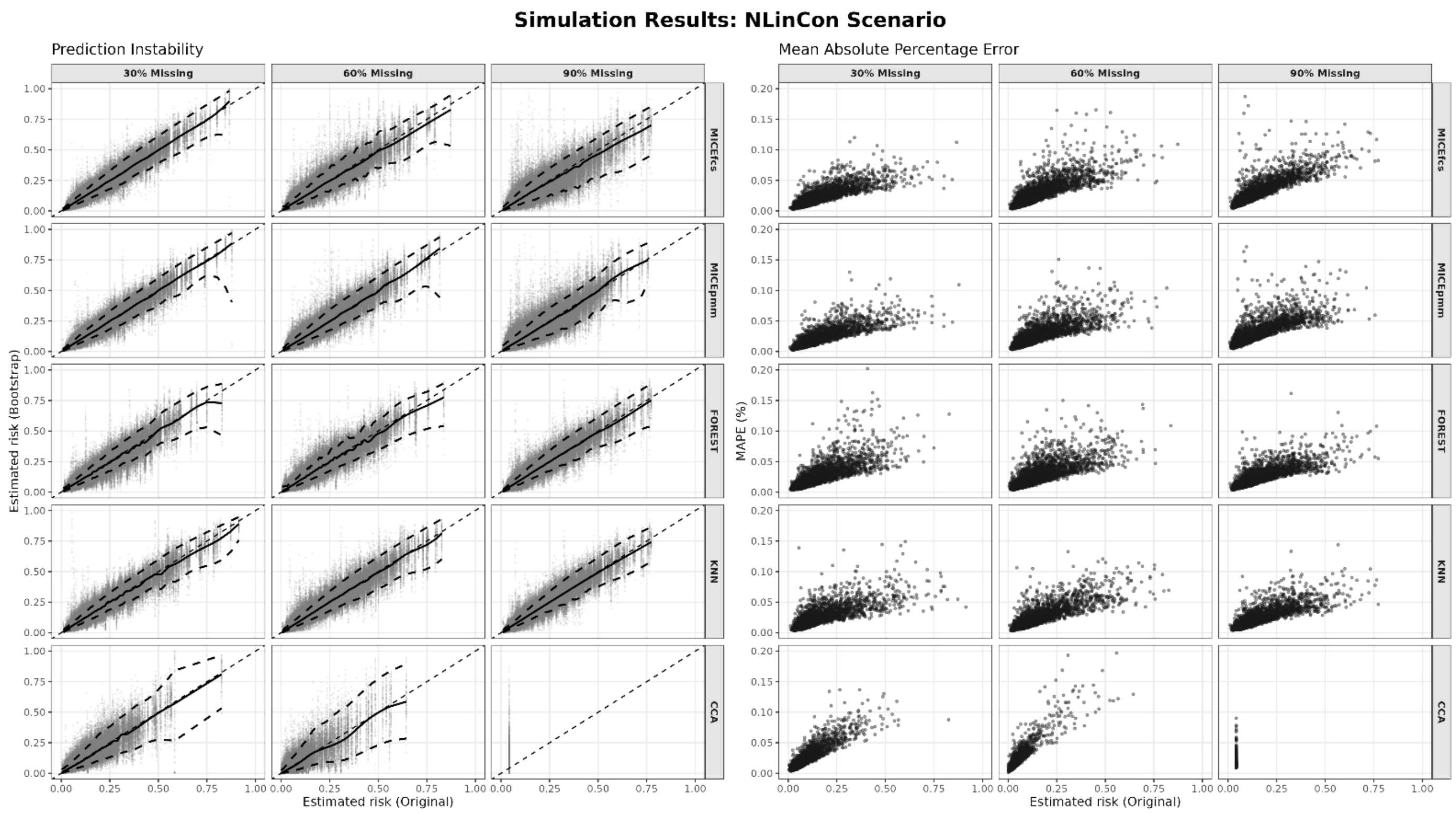

**Supplementary Figure S5.** Instability plot and mean absolute predictor error (MAPE) plot across all missingness proportions in missing of non-linear continuous and categorical (NLinMix) scenarios

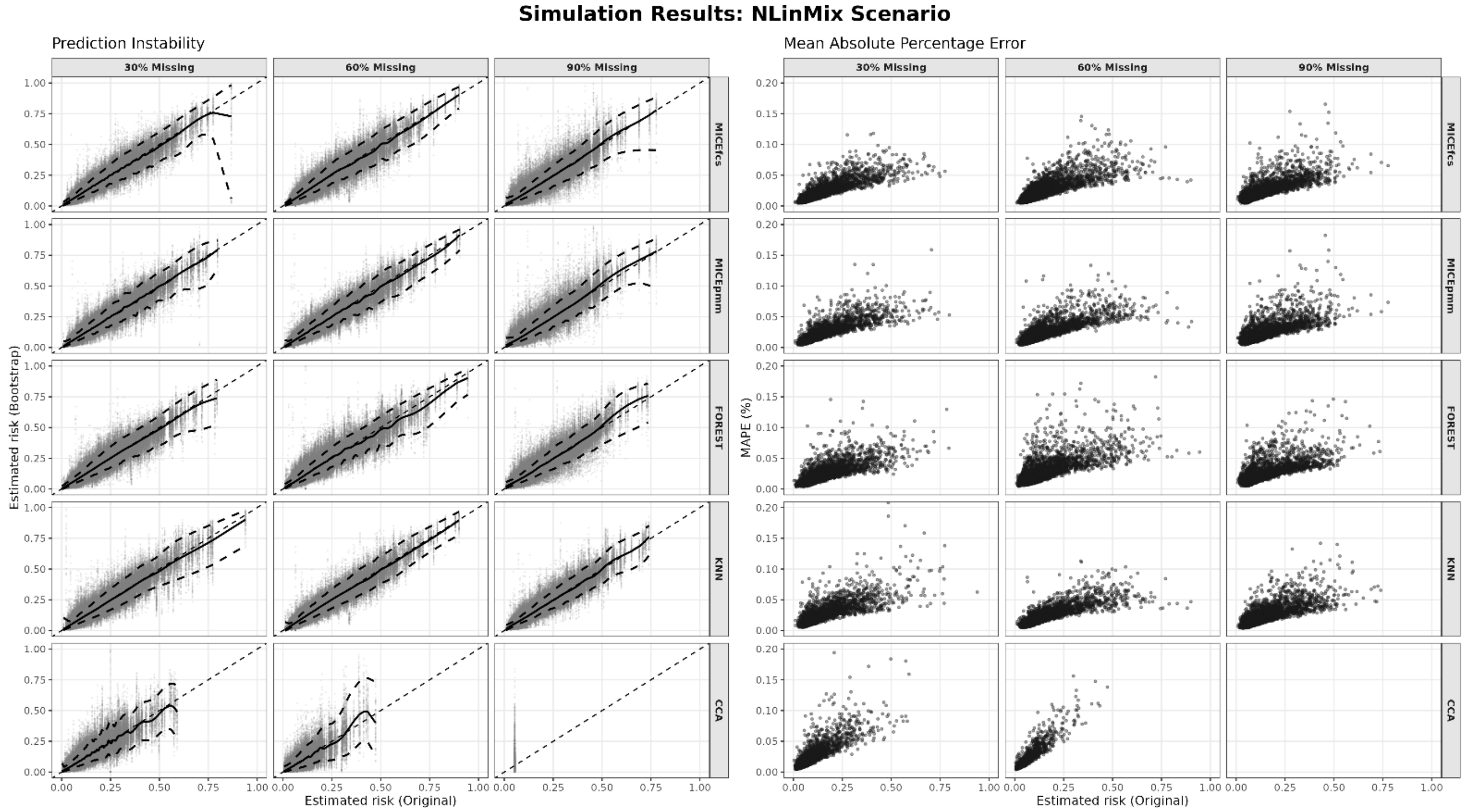

**Supplementary Figure S6.** Instability plot and mean absolute predictor error (MAPE) plot across all missingness proportions in missing of all condition's scenarios

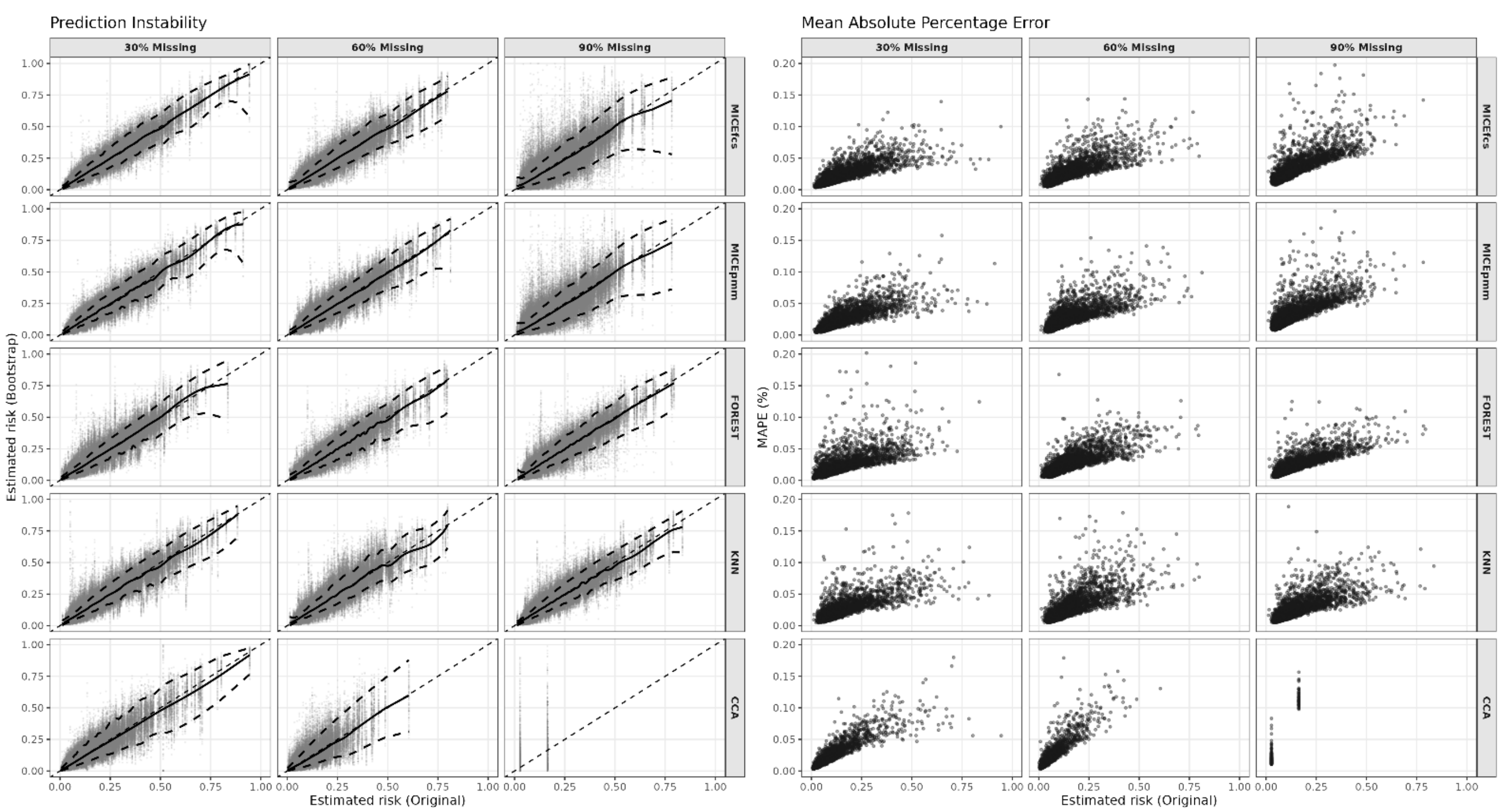

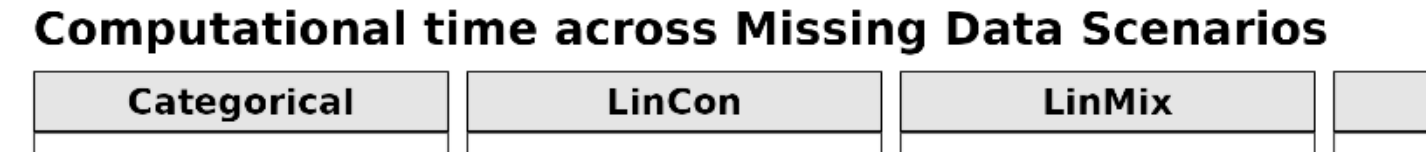
**Supplementary Figure S7.** Comparison of Computational time across all scenarios and imputation methods

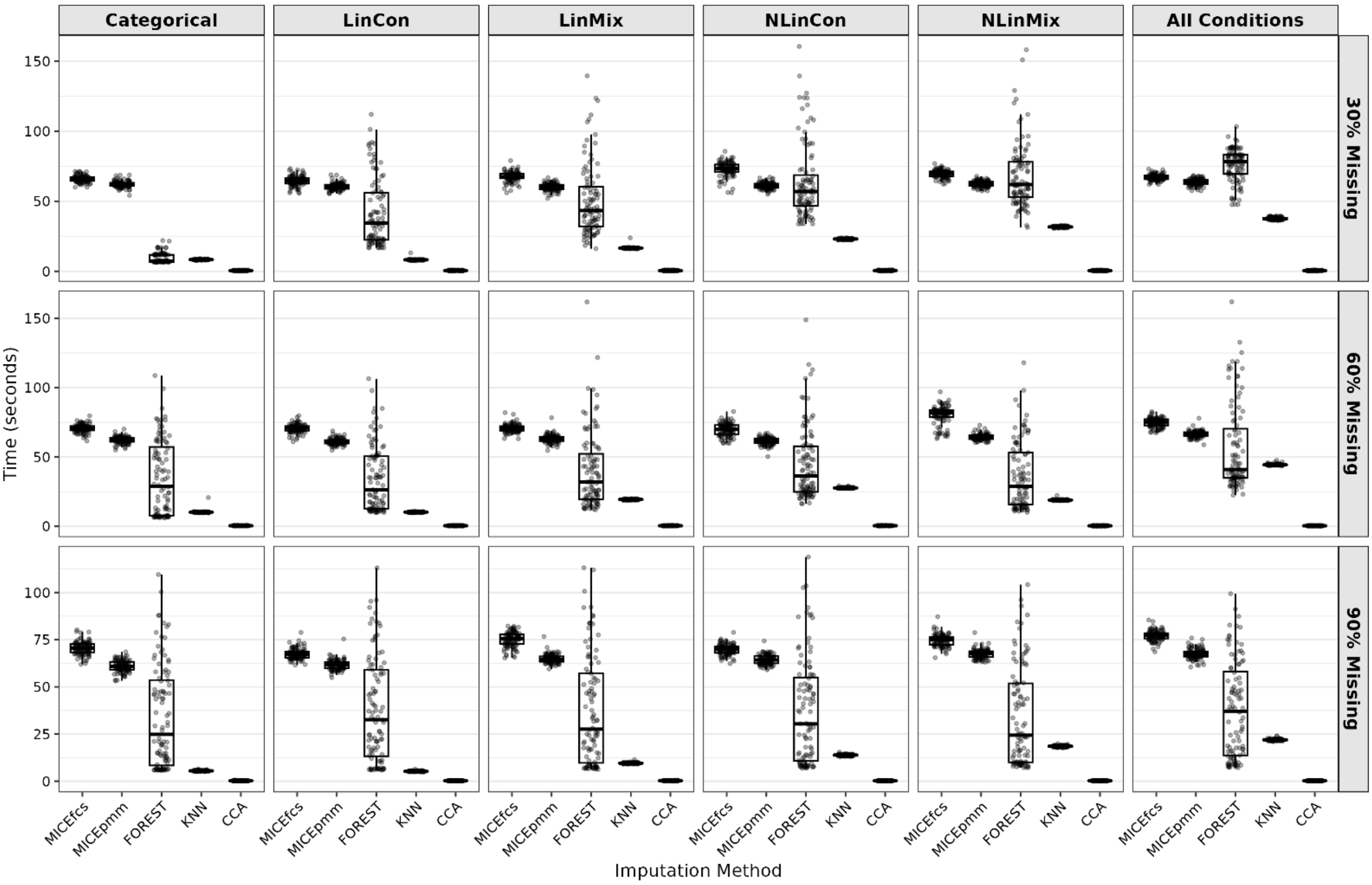